\documentclass[journal,onecolumn,12pt,draftclsnofoot]{IEEEtran}
\usepackage{amsmath}
\usepackage{amsthm}
\usepackage{amssymb}   
\usepackage{graphicx}
\usepackage{subfigure}
\usepackage{color}
\usepackage{balance}
\usepackage{cite}
\usepackage{url}
\usepackage{relsize}

\newcommand{\TDAoI}{\Delta^{\textsf{\!\smaller[3]{2D}}}}
\newcommand{\AbbrTDAoI}{2D-AoI}
\begin{document}
\title{\AbbrTDAoI{}: Age-of-Information of Distributed Sensors for Spatio-Temporal Processes}
\author{Markus Fidler, Flavio Gallistl, Jaya Prakash Champati, Joerg Widmer\thanks{M. Fidler and F. Gallistl are with the Department of Electrical Engineering and Computer Science, Leibniz University Hannover. J. P. Champati is with the University of Victoria. J. Widmer is with the IMDEA Networks Institute, Madrid. This work was supported in part by the German Research Foundation (DFG) under grant FI 1236/9.1.}}
\maketitle
\begin{abstract}
The freshness of sensor data is critical for all types of cyber-physical systems. An established measure for quantifying data freshness is the Age-of-Information (AoI), which has been the subject of extensive research. Recently, there has been increased interest in multi-sensor systems: redundant sensors producing samples of the same physical process, sensors such as cameras producing overlapping views, or distributed sensors producing correlated samples. When the information from a particular sensor is outdated, fresh samples from other correlated sensors can be helpful. To quantify the utility of distant but correlated samples, we put forth a two-dimensional (2D) model of AoI that takes into account the sensor distance in an age-equivalent representation. Since we define \AbbrTDAoI{} as equivalent to AoI, it can be readily linked to existing AoI research, especially on parallel systems. We consider physical phenomena modeled as spatio-temporal processes and derive the \AbbrTDAoI{} for different Gaussian correlation kernels. For a basic exponential product kernel, we find that spatial distance causes an additive offset of the AoI, while for other kernels the effects of spatial distance are more complex and vary with time. Using our methodology, we evaluate the \AbbrTDAoI{} of different spatial topologies and sensor densities.
\end{abstract}
%
%
\section{Introduction}
The Internet of Things (IoT) aims to achieve seamless integration between the physical and digital worlds. Physical devices, equipped with sensors and network connectivity, exchange information that is collected and processed in cyberspace. This information is used to gather data about physical phenomena, make decisions, augment reality, or control devices in the physical world through actuators. Examples of such cyber-physical systems include environmental monitoring, intelligent transportation, augmented reality, robotics, and networked feedback control. Common to these is the need to maintain fresh information about a physical phenomenon. An accepted measure for quantifying information freshness is the Age-of-Information (AoI). 

The concept of AoI was initially defined in the field of vehicular communications~\cite{kaul:ageofinformationvehicular} and motivated the investigation of AoI in queueing systems~\cite{kaul:ageofinformationqueue}. Today, the AoI is known for a variety of system models, see the recent surveys~\cite{yates:ageofinformationsurvey, kosta:ageofinformation}, including wireless channels~\cite{pan:hybridchannels, fidler:ageofinformationparallel}, wireless networks~\cite{modiano:informationfreshness}, random access channels~\cite{kaul:aoimultiaccess, pappas:ageofinformationnetworkcalculus, chen:aoirandomaccesschannels}, queueing systems~\cite{Soysal2021, inoue:aoisingleserverqueues}, and queueing networks~\cite{Bedewy2019:TON}. While the focus has been on mean AoI and mean peak AoI, there are a number of papers that derive the CCDF~\cite{inoue:aoisingleserverqueues, rizk:palmaoi, champati:ageofinformationgigiqueue} or tail bounds of the AoI~\cite{champati:ageofinformationmaxplus, noroozi:minplusaoi}. 

\begin{figure}
\subfigure[System with $S$ networked sensors]{\includegraphics[width=0.48\linewidth]{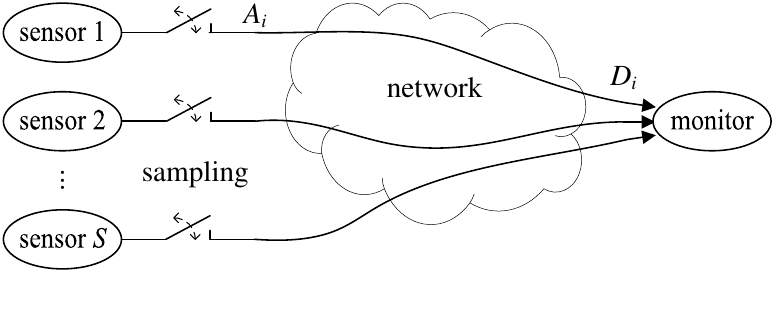}\label{fig:system}}
\hfill
\subfigure[Temporal and spatial prediction]{\includegraphics[width=0.48\linewidth]{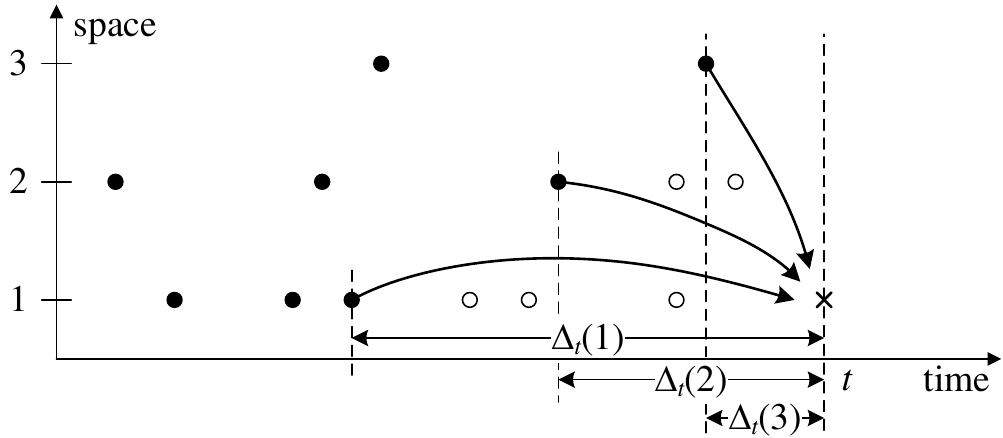}\label{fig:predictionexample}}
\caption{Fig.~\ref{fig:system}: System with $S=3$ sensors that observe a physical phenomenon. The sensors generate samples, indexed by $i$, at times $A_i$. The samples are transmitted via a network and received by the monitor at times $D_i$. Fig.~\ref{fig:predictionexample}: Example of a timeline. The times $A_i$ at which sensors take samples are indicated by circles, with the circle being filled, if the sample is available at the monitor at time $t$, i.e., $D_i \le t$. Assume the physical process at position 1 and time $t$ is of interest. A stale sample of sensor 1 with AoI $\Delta_t(1)$ can be used, or a prediction can be made from fresher samples from sensors 2 and 3.}
\end{figure}
The classic configuration studied in AoI involves a single sensor that observes a physical phenomenon and transmits sensor samples over a network to a monitor. The system is shown in Fig.~\ref{fig:system} as a special case for $S=1$ sensor. The sensor takes samples indexed $i \ge 1$ at times $A_i$ and sends the samples as packets over the network. The network can cause delays or packet loss and packets depart from the network to the monitor at times $D_i \ge A_i$, or $D_i=\infty$ if a packet cannot be recovered. 

In many instances of practical significance, however, there are multiple sensors, $S > 1$, as illustrated in Fig.~\ref{fig:system}. Sensors can be spatially distributed, redundant, or of different types, and can be connected in a variety of ways, including independent network paths or a shared wireless channel. Environmental monitoring involves many spatially distributed sensors that generate correlated information and can be connected via LoRaWAN or 5G mMTC, for example. Modern vehicles are equipped with automotive Ethernet and an array of sensors, including cameras, lidar, radar, and ultrasound, that observe the environment and create overlapping or correlated views. Vehicles share sensor information over Wifi to create a collective perception, to name a few examples. Despite this, the effects of the AoI when multiple statistically correlated sensors are available have only recently begun to be explored~\cite{hribar2018using, li2022age, hakansson2022schedulingspatiotemporalgaussian, jiang2019howdensely, modiano:aoicorrelatedsourceskalmanfilter}, and so far there is no mathematical derivation that incorporates spatial correlation into the AoI. 

In this study, we examine physical phenomena represented as correlated spatio-temporal processes sampled by a number of distributed sensors $S$. The spatial dimension can be Euclidean space, although this is not a necessary condition. The same methodology can be applied to include sensors that observe different but correlated physical phenomena. To illustrate, the spatial dimension could include road traffic volume and travel time or even air quality. In general, the spatial and temporal sampling densities are constrained by the limited capacity of the network. This raises the question of which sensor should be sampled and when. A related question is whether it is preferable to use outdated information obtained at a specific location or to utilize the most recent data from a different location. Fig.~\ref{fig:predictionexample} illustrates the example. To make progress on these issues, it is necessary to formalize the trade-off between AoI and the spatial distance of sensors.

To account for both temporal and spatial dimensions, this paper presents a two-dimensional (2D) model of AoI that uses an age-equivalent representation of spatial distance. The distance measure can be directly incorporated into the AoI, providing a strong link to the existing AoI literature and a foundation for further investigation. In particular, our work achieves a mapping of the AoI of spatio-temporal sampling to a suitably adapted model of AoI in parallel systems. The spatio-temporal process model we start from is generic and can be defined in different ways depending on the specific analysis being performed. In the course of this paper we use Gaussian processes, with distinct correlation kernels, but other models are also possible. In the basic case of exponential product kernels, the spatial distance between the sensors leads to an additive shift of the AoI, that is independent of time. In contrast, in the case of squared exponential and rational quadratic kernels, the impact of spatial distance on the AoI is observed to diminish over time. 

By employing our model of \AbbrTDAoI{}, we demonstrate the potential benefits and limitations of distributed sensors. In our analysis, we consider different network models, including both lossy ALOHA multi-access links and lossless M$\mid$M$\mid$1 queues. We evaluate the impact of spatial topology and sensor density on the AoI performance. We believe that our methodology also lends itself to the study of a range of other relevant systems, an example is included in the appendix.

The remainder of this work is structured as follows. We start with a discussion of related works in Sec.~\ref{sec:relatedwork}. In Sec.~\ref{sec:spatiotemporaldistance} we define the notion of \AbbrTDAoI{}. We show how the \AbbrTDAoI{} can be evaluated for a network model with independent channels in Sec.~\ref{sec:networkmodel}. In Sec.~\ref{sec:processmodel} we use Gaussian processes to characterize spatio-temporal physical phenomena and use these to instantiate the \AbbrTDAoI{} model. Sec.~\ref{sec:catalogeofkernels} considers relevant covariance kernels and their impact on the AoI. In Sec.~\ref{sec:evaluation} we show evaluation results for the density of a sensor network served by lossless, lossy, independent, or non-independent channels. Brief conclusions are provided in Sec.~\ref{sec:conclusions}. A background to Gaussian process regression, an example of a non-Gaussian model of \AbbrTDAoI{}, as well as some further results are given in the appendix.
%
%
\section{Related Work}
\label{sec:relatedwork}
The freshness of information is of general interest in various areas, including vehicular networks~\cite{kaul:ageofinformationvehicular, kaul:ageofinformationqueue}, remote state estimation~\cite{sun:remoteestimation, ornee:remoteestimation, modiano:aoicorrelatedsourceskalmanfilter}, networked feedback control~\cite{champati:ageofinformationfeedbackcontrol, klugel:aoipenalty, ayan:valueofinformation}, and cyber-physical systems in general, databases, caches, to micro-blogging~\cite{altman:ageofinformationmicroblogging}. The work~\cite{kaul:ageofinformationvehicular} on the AoI of status information broadcast by vehicles in a vehicular network marks the beginning of comprehensive research on the topic of AoI. Subsequently, the AoI of various queueing models was investigated~\cite{kaul:ageofinformationqueue}, and today a catalog of results for various systems is available, see e.g.,~\cite{inoue:aoisingleserverqueues} and the surveys~\cite{yates:ageofinformationsurvey, kosta:ageofinformation}.

For a basic definition of AoI, consider a sensor that is connected to a monitor via a network, as shown in Fig.~\ref{fig:system} for the special case $S=1$. The sensor generates samples $i \ge 1$ at times $A_i$ and sends them to the network, which delivers them to the monitor at times $D_i$. The AoI at time $t \ge D_1$ is defined as
\begin{equation}
\Delta_t = t - \max_{i\ge 1} \{A_i : D_i \le t\} .
\label{eq:agesimple}
\end{equation}
The AoI exhibits a distinctive piecewise linear sawtooth shape. Upon the arrival of a new sample at the monitor, the AoI drops to the delay of that sample, that is $\Delta_t = D_i - A_i$ if $t = D_i$. Subsequently, the AoI increases at unit rate with time, until a fresher sample is received. 

However, there are also systems that exhibit nonlinear behavior as the information ages. Important examples arise in remote state estimation of a system or a process~\cite{sun:remoteestimation, ornee:remoteestimation, champati:ageofinformationfeedbackcontrol, klugel:aoipenalty, ayan:valueofinformation}. The difference to Eq.~\eqref{eq:agesimple} is often formalized in terms of nonlinear functions of the AoI~\cite{yates:ageofinformationsurvey}. An age-penalty is defined in~\cite{shroff:updateorwait} as a non-decreasing function of the AoI to model the dissatisfaction caused by stale information. The concept of the age of incorrect information~\cite{maatouk:ageofincorrectinformation} uses a product of two functions, an age-penalty and an error-penalty, where the latter evaluates the deviation of the monitor's estimate from the correct state of the physical process. Mirroring age-penalty functions, the non-increasing value of aging information is modelled as value-of-information in~\cite{kosta:valueofinformation}.

Nonlinear functions of the AoI have been used in networked feedback control systems to characterize the mean squared error that arises in remote state estimation of a linear system with Gaussian disturbances~\cite{champati:ageofinformationfeedbackcontrol, klugel:aoipenalty, ayan:valueofinformation}. Related to this are the works~\cite{sun:remoteestimation, ornee:remoteestimation} on remote estimation of the state of Gaussian processes. Gaussian processes are also used to model spatial phenomena. In this case, the process at one location of interest can be predicted from another sensor location. The prediction variance is used in~\cite{krause:communitysensing} to define the value-of-information of that sensor. 

In recent research~\cite{yates:aoimultiplesources, sun:aoimultipleflows, javani:aoimultiplesensing, gindullina:aoisourcediversity, chen:agedualupdatingsystem, saad2020aoimaxofm, fodor2019ageoverlappingcameras, popovski2019multiplesensors, popovski2023multiplesensors, modiano:aoicorrelatedsources, tong2022aoischedulingcorrelatedsources, hribar2018using, li2022age, hakansson2022schedulingspatiotemporalgaussian, jiang2019howdensely, heinovski2023focusing, zancanaro2023voicorrelatedsources, modiano:aoicorrelatedsourceskalmanfilter, chiariotti:multisensorvoi2, pappas:multiplemarkovsources, crosara:singlequeuetwosource, crosara:singlequeuemanysource}, there is a growing interest in the AoI of multi-sensor systems, as shown as an example in Fig.~\ref{fig:system}. The works differ in their assumptions regarding the overlap or correlation of the sensors. In~\cite{yates:aoimultiplesources, sun:aoimultipleflows} multiple sensors that generate independent information flows are considered. The flows share a single queue, and thus influence each other's AoI. The same scenario, but with correlated sources, is considered in~\cite{crosara:singlequeuetwosource, crosara:singlequeuemanysource}, where a fraction of the updates of each source are also valid updates for the other sources.

The AoI of systems in which multiple sensors monitor the same physical phenomenon but transmit their samples via independent channels is studied in~\cite{gindullina:aoisourcediversity, chen:agedualupdatingsystem}. In~\cite{gindullina:aoisourcediversity}, an energy harvesting system is investigated that can choose between multiple sensors, each of which offers an individual trade-off between AoI and energy. A dual-updating system, where two sensors generate and transmit samples independently, is considered in~\cite{chen:agedualupdatingsystem}, and the AoI is derived for Markovian and deterministic queues with feedback. Directly related to multi-sensor systems that observe a single physical phenomenon are parallel systems in which the samples from a single sensor are split between two (or more) independent channels~\cite{bhati:parallelmm1age, fidler:ageofinformationparallel}. As only the most recent sample is decisive for the AoI, the main characteristic of parallel systems is that their AoI is the minimum of the AoI of the individual subsystems. In contrast,~\cite{saad2020aoimaxofm} studies a system with $S$ sensors with the constraint that a sample from each of the sensors is required to reconstruct the physical process. In this system, the AoI is the maximum of the AoI of the individual sensors.

The AoI of systems with overlapping sensors that generate common observations is studied in~\cite{fodor2019ageoverlappingcameras, javani:aoimultiplesensing, popovski2019multiplesensors, popovski2023multiplesensors, modiano:aoicorrelatedsources, tong2022aoischedulingcorrelatedsources}. In~\cite{fodor2019ageoverlappingcameras} a multi-view camera system is considered, where $S$ cameras observe $K$ different scenes. A single camera can observe multiple scenes and the cameras' fields of view overlap so that scenes are observed by multiple cameras simultaneously. A similar model of sensors with overlapping monitoring areas is used in~\cite{tong2022aoischedulingcorrelatedsources} and in~\cite{javani:aoimultiplesensing}, where $K$ information sources are sensed by $S$ servers. A related model with non-deterministic sensors, where sensor $s \in \{1,2,\dots,S\}$ detects physical phenomenon $k \in \{1,2,\dots,K\}$ with probability $p_{sk}$, is studied in~\cite{popovski2019multiplesensors, popovski2023multiplesensors} and also in~\cite{modiano:aoicorrelatedsources} which models the coupling of the AoI processes of the sources. 

More closely related to our work are the recent papers~\cite{hribar2018using, li2022age, hakansson2022schedulingspatiotemporalgaussian, jiang2019howdensely} that use Gaussian processes to model the spatial and temporal correlations of multiple distributed sensors. A common feature of these papers is that the monitor uses only the most recent sample from each sensor. This is consistent with the definition of AoI. The AoI is then used as input to a nonlinear function that expresses an age-penalty or value-of-information: \cite{hribar2018using} employs the estimation error resulting from using aged information from a possibly remote sensor, similarly~\cite{hakansson2022schedulingspatiotemporalgaussian, jiang2019howdensely} apply the mean squared estimation error, and~\cite{li2022age} uses the mutual information that applies over a spatial span and temporal lag. A related work is also~\cite{modiano:aoicorrelatedsourceskalmanfilter}, which uses Kalman filters with inputs from multiple sensors to estimate the state of several random walks with spatially correlated Gaussian increments. Unlike~\cite{hribar2018using, li2022age, hakansson2022schedulingspatiotemporalgaussian, jiang2019howdensely}, our consideration of spatially distributed sensors takes place in the AoI domain, i.e., we transform spatial distance into a representation that is equivalent to AoI. This provides a direct link to the AoI literature, where our result generalizes the AoI of parallel systems~\cite{bhati:parallelmm1age, fidler:ageofinformationparallel} by considering correlated spatio-temporal processes instead of a single temporal process.

In more detail~\cite{hribar2018using} considers two sensor nodes, one at the point of interest and one distant. The sensors generate samples periodically with update period $T$ and a defined offset $\tau$. The AoI is deterministic in $[0,T)$ meaning there are no network delays or losses. Regarding the estimation error, a key insight is that there is a break-even point at which a fresh sample from the far sensor is better than an outdated sample from the near sensor. This imposes a deterministic condition on the update period and distance for the far sensor to take effect. In our work, we show that similar conditions can be established in the case of random network delays as they occur, e.g., in multi-access channels or queueing networks. 

The papers~\cite{li2022age, hakansson2022schedulingspatiotemporalgaussian} differ from our work in that they consider the optimal scheduling of $S$ sensors connected via a satellite link with deterministic delay or a wireless sensor network with zero delay, respectively. For a random topology where sensors are deployed as a Poisson point process,~\cite{jiang2019howdensely} investigates how the density of exponential temporal and spatial sampling affects the estimation error. Analytical solutions are provided for a one-dimensional sensor network and exponential service times. Among other things, we also evaluate the density of a sensor grid, but our results are expressed in terms of AoI.

Unlike~\cite{hribar2018using, li2022age, hakansson2022schedulingspatiotemporalgaussian, jiang2019howdensely}, the works~\cite{tong2022aoischedulingcorrelatedsources, heinovski2023focusing, zancanaro2023voicorrelatedsources}, like ours, transfer the spatial distance of sensors into the AoI domain: \cite{tong2022aoischedulingcorrelatedsources} multiplies the AoI of a sensor by an attenuation factor $a \in [0,1]$ when a more recent sample from another correlated sensor becomes available. Similarly,~\cite{zancanaro2023voicorrelatedsources} defines a hyper-parameter $a$ as a weight of the AoI of a neighbor having fresher information, and~\cite{heinovski2023focusing} uses a weighting factor $a$ of the AoI, which is determined from a spatial model of a vehicular network consisting of spatial distance and angle of vehicles. In contrast to these works, we do not make the a priori assumption that spatial distance can be expressed by a constant attenuation or weighting factor of the AoI. Instead, we start with a defined correlation model and derive a function that characterizes the influence of spatial distance on the AoI. To parameterize the function, we evaluate several spatio-temporal correlation kernels. Stationary exponential product kernels, which are also used in~\cite{hribar2018using, li2022age, jiang2019howdensely},
have the basic effect of causing a constant additive offset of the AoI of distant sensors. For squared exponential and rational quadratic kernels, on the other hand, we find that the influence of spatial distance on the AoI diminishes as the AoI increases, which is a fundamental difference of the models.
%
%
\section{Definition of \AbbrTDAoI{}}
\label{sec:spatiotemporaldistance}
We consider a set of sensors $\mathbb{S} = \{1,2,\dots,S\}$. In a basic scenario, sensors are homogeneous and differ only by their position $x_s$ where $s \in \mathbb{S}$ is the index of the sensor. The position $x_s$ can be scalar, i.e. lie on a one-dimensional line, but it can also lie in two- or three-dimensional Euclidean space. Different types of sensors create an additional dimension that can be included in a higher dimensional model of $x_s$. This can also be handled using the method described below. The sensors sample the physical process and transmit the samples to a monitor. The samples may be subject to noise. We will consider noise in Sec.~\ref{sec:observationnoise}. During transmission, samples may be delayed or may even get lost. We denote samples that are available at the monitor at time $t \ge 0$ by indicator variables $\Gamma_t(s,\tau) \in \{0,1\}$, where $s \in \mathbb{S}$ and $\tau \in [0,t]$ is the generation time of the sample. Time can be continuous or discrete. The monitor can have an unlimited or a limited time horizon $T$, in which case samples that are older than $T$ are discarded. The AoI of sensor $s \in \mathbb{S}$ at time $t \ge 0$ is
\begin{equation}
\Delta_t(s) = \inf \{\tau \in [0,\min\{t,T\}] : \Gamma_t(s,t-\tau) = 1 \}.    
\label{eq:aoifromindicators}
\end{equation}
If no sample of sensor $s$ is available at the monitor at time $t$, we define $\Delta_t(s) = \infty$, that is also obtained from Eq.~\eqref{eq:aoifromindicators} as the result of taking the infimum of the empty set. 

The AoI $\Delta_t(s)$ can be thought of as a penalty when predicting the physical process at position $x_s$ and time $t$ from the latest sample that is available for that position at the monitor. This is the sample generated at time $\tau = t-\Delta_t(s)$. The actual quality of the prediction depends on the temporal correlation.

If the point of interest is $x_s$, but we have a sensor at position $x_{\varsigma} \neq x_s$ (instead of or in addition to the sensor at position $x_s$), the physical process at position $x_s$ and at time $t$ can also be predicted from a sample of the sensor at position $x_{\varsigma}$. The latest sample of that sensor was generated at time $\tau = t-\Delta_t(\varsigma)$. Now, the quality of the prediction depends on the temporal correlation and on the spatial correlation of the process at positions $x_{\varsigma}$ and $x_s$. This motivates a notion of two-dimensional (2D) AoI, expressed by the function $\TDAoI_t(\varsigma,s)$. The \AbbrTDAoI{} can be thought of as a spatio-temporal distance of a sample. In the special case $\varsigma = s$ we have $\TDAoI_t(\varsigma,\varsigma) = \Delta_t(\varsigma)$, while for $\varsigma \neq s$ in the general case $\TDAoI_t(\varsigma,s)$ is different from $\Delta_t(\varsigma)$. 

The difference $\Lambda_t(\varsigma, s) = \TDAoI_t(\varsigma, s) - \Delta_t(\varsigma)$ represents the effect of the distance of the sensor at position $x_{\varsigma}$ on the AoI at position $x_s$. We call $\Lambda_t(\varsigma,s)$ the Age-equivalent (of the) Distance (AeD). In the simplest case, a distant sensor perceives a physical phenomenon delayed by a deterministic time $\Lambda_t(\varsigma,s)$. This is the case with seismic waves, for example, which travel at a few kilometres per second and are detected by distributed sensors. How to select $\Lambda_t(\varsigma,s)$ in non-trivial cases based on statistical correlation is shown in the following Secs.~\ref{sec:processmodel} and~\ref{sec:catalogeofkernels}. In some relevant cases, the AeD $\Lambda_t(\varsigma,s) = \Lambda(\varsigma,s)$ is only a function of space and independent of time, while in others it is a more complex function that changes with time. 

\begin{figure}
\centering
\includegraphics[width=0.5\linewidth]{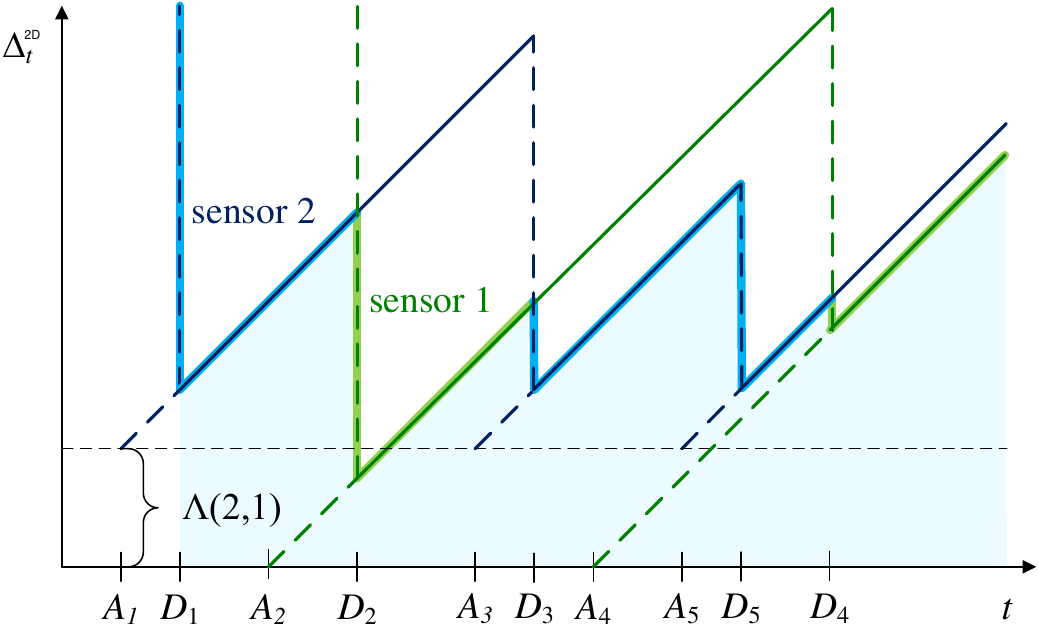}    
\caption{Example of the evolution of the \AbbrTDAoI{} $\TDAoI_t(1)$ for $S=2$ sensors.}
\label{fig:spatiotemporaldistance}
\end{figure}

Using the definition of AeD, the \AbbrTDAoI{} is expressed as a sum of AoI and AeD as
\begin{equation}   
\TDAoI_t(\varsigma,s) = \Delta_t(\varsigma) + \Lambda_t(\varsigma,s) .
\label{eq:spatialdistance}
\end{equation}
Given the monitor has samples from several sensors $S > 1$, an evident choice is to select the sample that has the minimal \AbbrTDAoI{}
\begin{equation}
\TDAoI_{t}(s) = \min_{\varsigma \in \mathbb{S}} \{\TDAoI_t(\varsigma,s)\} . 
\label{eq:minimalspatiotemporaldistance}
\end{equation}
To keep the notation simple, we use $\TDAoI_{t}$ again, but with a single argument. In the case that the AeD is just a function of space, it simply contributes as a time-independent offset, and we have
\begin{equation}
\TDAoI_t(s) = \min_{\varsigma \in \mathbb{S}} \{\Lambda(\varsigma,s) + \Delta_t(\varsigma)\} .
\label{eq:minimalspatiotemporaldistanceoffset}
\end{equation}
Eq.~\eqref{eq:minimalspatiotemporaldistanceoffset} permits a very intuitive visualization, which is shown in Fig.~\ref{fig:spatiotemporaldistance}. In the example, there are $S=2$ sensors and the minimal \AbbrTDAoI{} $\TDAoI_t(1)$ at position $1$ is derived. As before, $A_i$ denotes the time of generation of sample $i$, that is the arrival time of the sample to the network, and $D_i$ the time of its departure from the network to the monitor. Samples 2 and 4 are taken by sensor 1 and samples 1, 3, and 5 by sensor 2. The thin green saw-tooth function represent $\TDAoI_t(1,1) = \Delta_t(1)$, that is the AoI of sensor 1. The thin blue saw-tooth represents $\TDAoI_t(2,1) = \Lambda(2,1) + \Delta_t(2)$, that is the AoI of sensor 2 shifted upwards by an offset that considers the AeD of sensor 2 with respect to position 1. At the monitor the sample that has the minimal \AbbrTDAoI{} is selected, marked by the thick saw-tooth function that is either green, if the most recent sample of sensor 1 is used, or blue, if the most recent sample of sensor 2 is used. 

It is evident that the utility of sensor 2 depends on the condition that the AoI of sensor 1 exceeds the AeD of sensor 2. In a deterministic network with constant delay, it is straightforward to devise a sampling schedule that adheres to the specified condition. This result is consistent with the identification of a break-even point in Gaussian process estimation from two sensors and deterministic sampling as detailed in~\cite{hribar2018using}, see the discussion in Sec.~\ref{sec:relatedwork}. In the following, we derive statistics on the \AbbrTDAoI{} under random sampling and random transmission delays.
%
%
\section{\AbbrTDAoI{} for Independent Channels}
\label{sec:networkmodel}
We consider a network that connects each sensor $\varsigma \in \mathbb{S}$ to the monitor via an independent channel. In this case, the distribution of the minimal \AbbrTDAoI{} from Eq.~\eqref{eq:minimalspatiotemporaldistance} is
\begin{align}
\mathsf{P} [\TDAoI_t(s) > y] &= \mathsf{P} \Bigl[\min_{\varsigma \in \mathbb{S}} \{\TDAoI_t(\varsigma,s)\} >  y\Bigr] \nonumber \\
&= \prod_{\varsigma \in \mathbb{S}} \mathsf{P} [\TDAoI_t(\varsigma,s) > y ] \label{eq:spatiotemporaldistanceccdfgeneral} \\
&= \prod_{\varsigma \in \mathbb{S}} \mathsf{P} [\Delta_t(\varsigma) > y - \Lambda(\varsigma,s)] . \label{eq:spatiotemporaldistanceccdflinear}
\end{align}
In Eq.~\eqref{eq:spatiotemporaldistanceccdflinear} we have substituted Eq.~\eqref{eq:spatialdistance} where we consider the AeD $\Lambda(\varsigma,s)$ as a deterministic, time-independent function of space and $\Delta_t(\varsigma)$ is the random AoI process. Note that the AeD can also be modeled as random, in which case Eq.~\eqref{eq:spatiotemporaldistanceccdflinear} evaluates the sum of two random variables $\mathsf{P}[\Delta_t(\varsigma)+\Lambda(\varsigma,s) > y]$, see Appendix~\ref{appendix:random_AeD}. For the special case $\Lambda(\varsigma,s) = 0$ for all $\varsigma,s$, Eq.~\eqref{eq:spatiotemporaldistanceccdflinear} recovers the AoI of a parallel system, in which independent samples of the same physical process are transmitted to a monitor via parallel transmission channels. Parallel systems have previously been studied in~\cite{bhati:parallelmm1age, fidler:ageofinformationparallel}. To evaluate Eq.~\eqref{eq:spatiotemporaldistanceccdflinear}, we need the CCDF of the AoI, as given in~\cite{inoue:aoisingleserverqueues} for the M$\mid$M$\mid$1 queue, for example.

In the following, we investigate spatio-temporal processes where $\Lambda(\varsigma,s) > 0$ for $\varsigma \neq s$. We consider the stationary distribution of $\TDAoI_t(s)$ and omit the subscript $t$ when it is nonambiguous. The expected value of the \AbbrTDAoI{} is obtained by integration of the tail distribution as
\begin{equation}
\mathsf{E}[\TDAoI(s)] = \int_0^\infty \mathsf{P} [\TDAoI(s) > y] dy .
\label{eq:meanspatiotemporalpenalty}
\end{equation}
%
%
\subsection{Sensors Connected via Independent M$\mid$M$\mid$1 Queues}
\label{sec:numericalexample1}
For a first evaluation, we consider a set of spatially distributed sensors that are each connected to the monitor via an independent M$\mid$M$\mid$1 queue. The CCDF of the AoI of the M$\mid$M$\mid$1 queue is known to be~\cite{inoue:aoisingleserverqueues}
\begin{equation}
\mathsf{P}[\Delta > y] = e^{-(1-\rho) \mu y} - \biggl(\frac{1}{1-\rho} + \rho \mu  y \biggr) e^{-\mu y} + \frac{1}{1-\rho} e^{-\lambda y},
\label{eq:ageccdfmm1}
\end{equation}
where parameter $y \ge 0$, $\lambda > 0$ is the arrival rate, $\mu > 0$ is the service rate, with $\mu > \lambda$ for stability, and $\rho = \lambda/\mu$ is the utilization. By insertion of Eq.~\eqref{eq:ageccdfmm1} into Eq.~\eqref{eq:spatiotemporaldistanceccdflinear}, we obtain the CCDF of the \AbbrTDAoI{} $\TDAoI(s)$. By integration of the tail distribution Eq.~\eqref{eq:meanspatiotemporalpenalty} the expected value $\mathsf{E}[\TDAoI(s)]$ follows. We note that the CCDF of the AoI or tail bounds of the AoI are also known for a number of other queueing systems and wireless channel models~\cite{inoue:aoisingleserverqueues, rizk:palmaoi, champati:ageofinformationgigiqueue, champati:ageofinformationmaxplus, noroozi:minplusaoi}, which can be used in the same way. 

\begin{figure}
\subfigure[Mean and 0.9 quantile]{\includegraphics[width=0.33\linewidth]{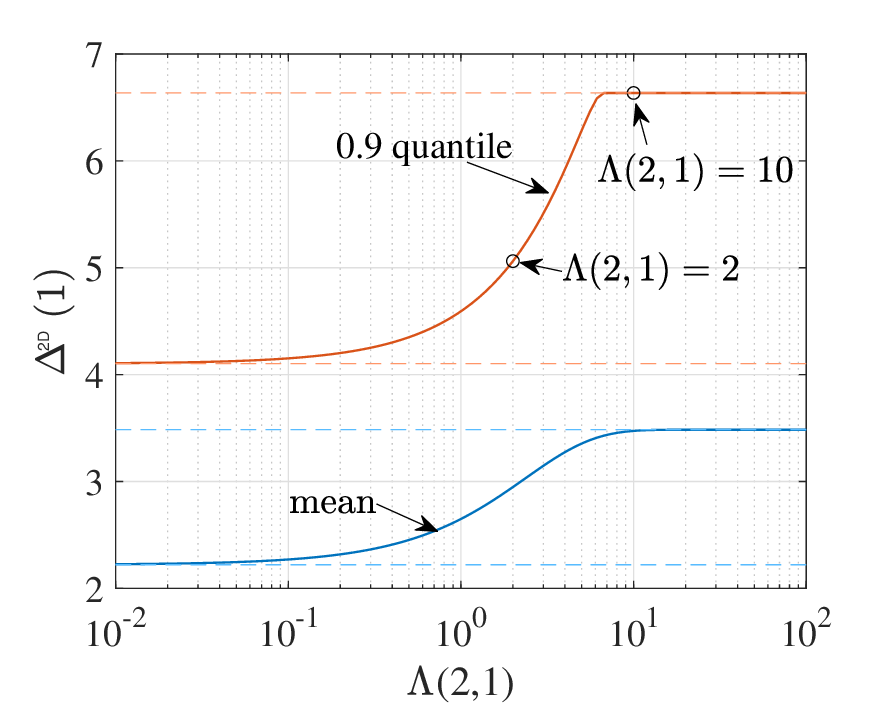}\label{fig:mm1spacevspenaltywithquantiles}}
\hfill
\subfigure[CCDF]{\includegraphics[width=0.33\linewidth]{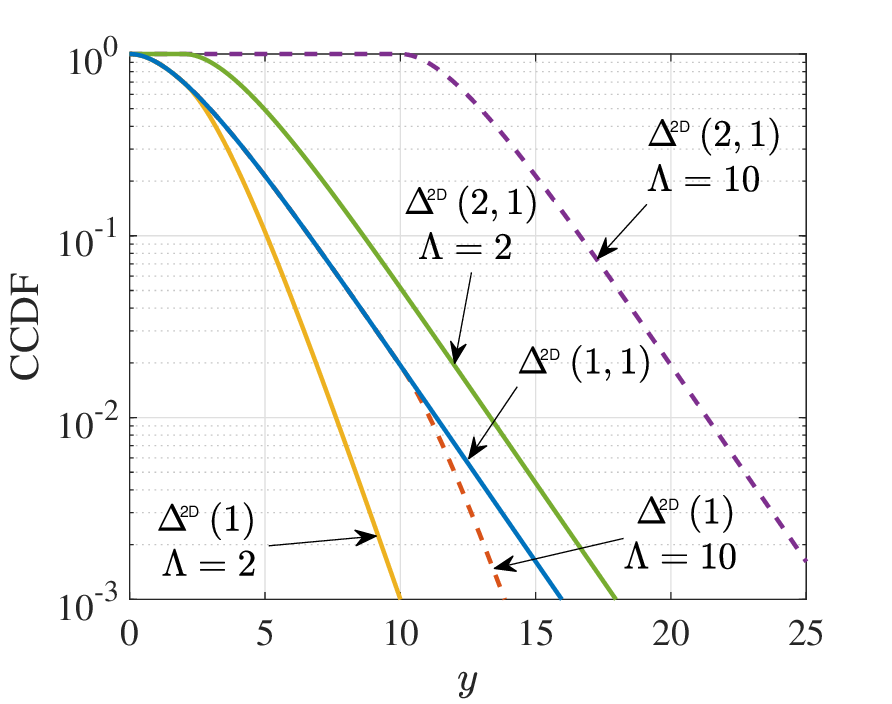}\label{fig:mm1CCDFsAll}}
\hfill
\subfigure[Allocation of service rates]{\includegraphics[width=0.31\linewidth]{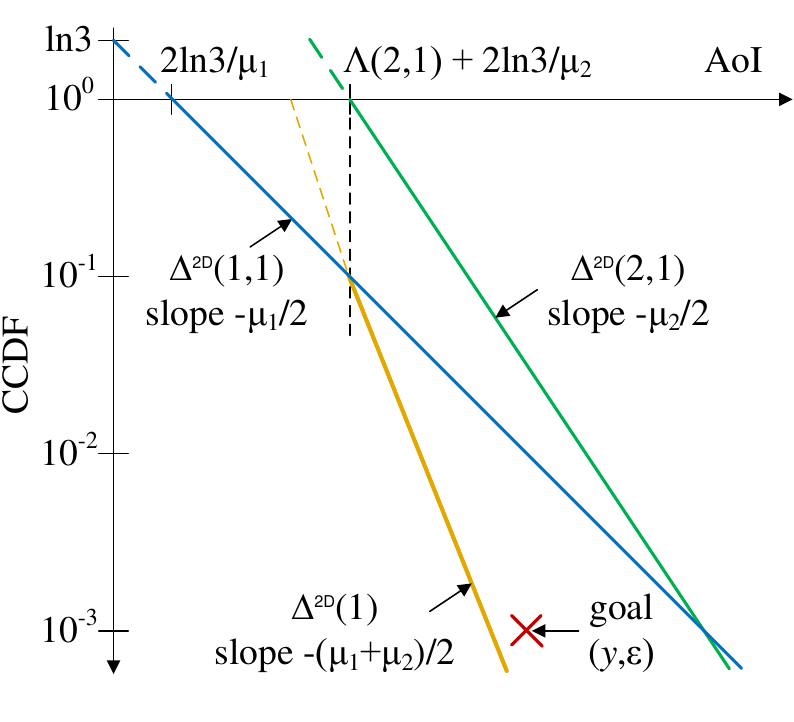}\label{fig:serviceprovisioning}}
\caption{\AbbrTDAoI{} $\TDAoI(\varsigma,s)$ and minimal \AbbrTDAoI{} $\TDAoI(s) = \min_{\varsigma} \{\TDAoI(\varsigma,s)\}$ for two sensors $\varsigma,s \in \{1,2\}$ that are connected to a monitor via independent M$\mid$M$\mid$1 queues. Fig.~\ref{fig:mm1spacevspenaltywithquantiles} evaluates the impact of the AeD $\Lambda(2,1)$ on $\TDAoI(1)$. Fig.~\ref{fig:mm1CCDFsAll} shows the effect of two values $\Lambda(2,1) \in \{2,10\}$ on the CCDFs. Fig.~\ref{fig:serviceprovisioning} illustrates how service rates can be allocated to achieve a target \AbbrTDAoI{} threshold of $y$ at $\varepsilon = 10^{-3}$. When individually considered $\TDAoI(1,1)$ of sensor 1 (blue line) and $\TDAoI(2,1)$ of sensor 2 (green line) are unable to achieve the goal and higher service rates $\mu_i$ would be needed. However, $\TDAoI(1)$ obtained by combining both sensors (yellow line) meets the goal.}
\label{fig:mm1CCDFs}
\end{figure}

In Fig.~\ref{fig:mm1CCDFs}, we consider $S=2$ sensors and evaluate the effect of the AeD $\Lambda(2,1)$ on the \AbbrTDAoI{} $\TDAoI(1)$. The M$\mid$M$\mid$1 queues are homogeneous with parameters $\lambda = 0.53$ and $\mu=1$. The utilization follows as $\rho = 0.53$, that is known to be optimal in the sense that it achieves the minimal expected AoI of a single M$\mid$M$\mid$1 queue~\cite{kaul:ageofinformationqueue}. 

In Fig.~\ref{fig:mm1spacevspenaltywithquantiles}, we show how the mean and the 0.9 quantile of $\TDAoI(1)$ increase with $\Lambda(2,1)$. In case of $\Lambda(2,1) = 0$, both sensors generate samples of the same process and $\TDAoI(1)$ equals the AoI of a basic parallel system, in which samples of a single sensor are transmitted in parallel via two independent M$\mid$M$\mid$1 queues. For $\Lambda(2,1) \rightarrow \infty$ the samples of sensor $2$ become irrelevant and $\TDAoI(1)$ approaches the AoI of a single M$\mid$M$\mid$1 queue. The cases $\Lambda(2,1) = 0$ and $\Lambda(2,1) \rightarrow \infty$ mark lower and upper bounds, which are indicated in the figure by dashed lines. In between, $\Lambda(2,1)$ has a noticeable effect. The extent is determined by $\Lambda(2,1)$ and by the speed of the tail decay of $\Delta_t(\varsigma)$ as given by Eq.~\eqref{eq:ageccdfmm1}. 

We explain the effect in Fig.~\ref{fig:mm1CCDFsAll} where CCDFs for the cases $\Lambda(2,1) = 2$ and $\Lambda(2,1) = 10$, marked by circles in Fig.~\ref{fig:mm1spacevspenaltywithquantiles}, are compared. For $\Lambda(2,1) = 2$ (solid lines), the CCDF of $\TDAoI(2,1)$ decays from 1.0 starting at $y = 2$, where the CCDF of $\TDAoI(1,1)$ has already dropped to approximately 0.6. At this point, sensor 2 begins to contribute and the CCDF of the minimal \AbbrTDAoI{} $\TDAoI(1)$ decays faster than $\TDAoI(1,1)$ from that point onward. For $\Lambda(2,1) = 10$ (dashed lines) the same effect occurs. However, the CCDF of $\TDAoI(1,1)$ has already dropped to about 0.02 for $y = 10$ before sensor 2 contributes, so that the benefit of sensor 2 is almost negligible. 

Essentially, the utility of sensor 2 depends on how far the CCDF of the AoI of sensor 1 has dropped before the AeD of sensor 2 is compensated and sensor 2 becomes useful. We note that the speed of the tail decay of the CCDFs depends on the arrival and service rates $\lambda$ and $\mu$ of the M$\mid$M$\mid$1 queues, where we used $\lambda = 0.53 \, \mu$, see the exponents in Eq.~\eqref{eq:ageccdfmm1}. 

We assemble a quick method that allows the service parameters to be set so that the \AbbrTDAoI{} exceeds a threshold value $y$ at most with a small probability $\varepsilon$. To do this, we estimate the tail of Eq.~\eqref{eq:ageccdfmm1} for $y \ge 2 \ln 3/\mu$ as
\begin{equation}
\mathsf{P}[\Delta > y] \le 3 e^{-\frac{\mu}{2} y} ,
\label{eq:ageccdfmm1quickestimate}
\end{equation}
where we used $\lambda  = 0.5 \mu$ and kept only the terms that have the dominant tail decay rate. For two sensors with service rates $\mu_1, \mu_2$, it follows by insertion of Eq.~\eqref{eq:ageccdfmm1quickestimate} into Eq.~\eqref{eq:spatiotemporaldistanceccdflinear} for $y \ge 2 \ln 3/\mu_1$ and $y \ge \Lambda(2,1) + 2 \ln 3/\mu_2$ that
\begin{align}
\mathsf{P}[\TDAoI(1) > y] & = \mathsf{P}[\Delta(1) > y] \; \mathsf{P}[\Delta(2) > y - \Lambda(2,1)] \nonumber \\
& \le 3 e^{-\frac{\mu_1}{2} y} \; 3 e^{-\frac{\mu_2}{2} (y - \Lambda(2,1))} \label{eq:agehdccdfmm1quickestimatefactors} \\
& = 9 e^{\frac{\mu_2}{2} \Lambda(2,1) } e^{-\frac{\mu_1+\mu_2}{2} y} =: \varepsilon .
\label{eq:agehdccdfmm1quickestimate}
\end{align}

Fig.~\ref{fig:serviceprovisioning} illustrates how the two factors of Eq.~\eqref{eq:agehdccdfmm1quickestimatefactors} contribute to $\TDAoI(1)$ and how the service rate of sensor 1 can be adjusted in order to attain a desired performance objective. The scenario is as follows: Given a position of interest 1, a target AoI threshold $y$ with excess probability $\varepsilon$, and a distant sensor 2 with AeD $\Lambda(2,1)$ and service rate $\mu_2$. Assume sensor 2 is close enough to position 1 to make a contribution, i.e., $\Lambda(2,1) < y - 2 \ln 3/\mu_2$, but it cannot guarantee the AoI threshold alone, i.e., $\mu_2 < 2 (\ln 3 - \ln \varepsilon)/(y-\Lambda(2,1))$. Hence, another sensor with index 1 is installed at the position of interest. The service rate $\mu_1$ of that sensor, which is required to fulfill the AoI threshold, is obtained by solving Eq.~\eqref{eq:agehdccdfmm1quickestimate} for
\begin{equation*}
\mu_1 = \frac{2(\ln 9 - \ln \varepsilon)}{y} - \mu_2 \biggl(1 - \frac{\Lambda(2,1)}{y} \biggr) .
\end{equation*}
A linear dependence of $\mu_1$ on $\mu_2$ is observed, where $\mu_2$ is discounted by a factor determined by the quotient of the AeD $\Lambda(2,1)$ and the AoI threshold $y$. If the spatial distance is zero, then $\Lambda(2,1) = 0$ and the system is a basic parallel system in which the service rate $\mu_2$ contributes without deduction. With increasing distance (larger $\Lambda(2,1)$) and more stringent AoI threshold (smaller $y$), however, the usefulness of sensor 2 diminishes until it is no longer able to contribute. In following results we also evaluate the impact of the service rate $\mu$ in larger topologies. 
%
%
\subsection{Resource Allocation in a Star Topology}
\label{sec:numericalexample2}
\begin{figure*}
\subfigure[Star topology]{\includegraphics[width=0.21\linewidth]{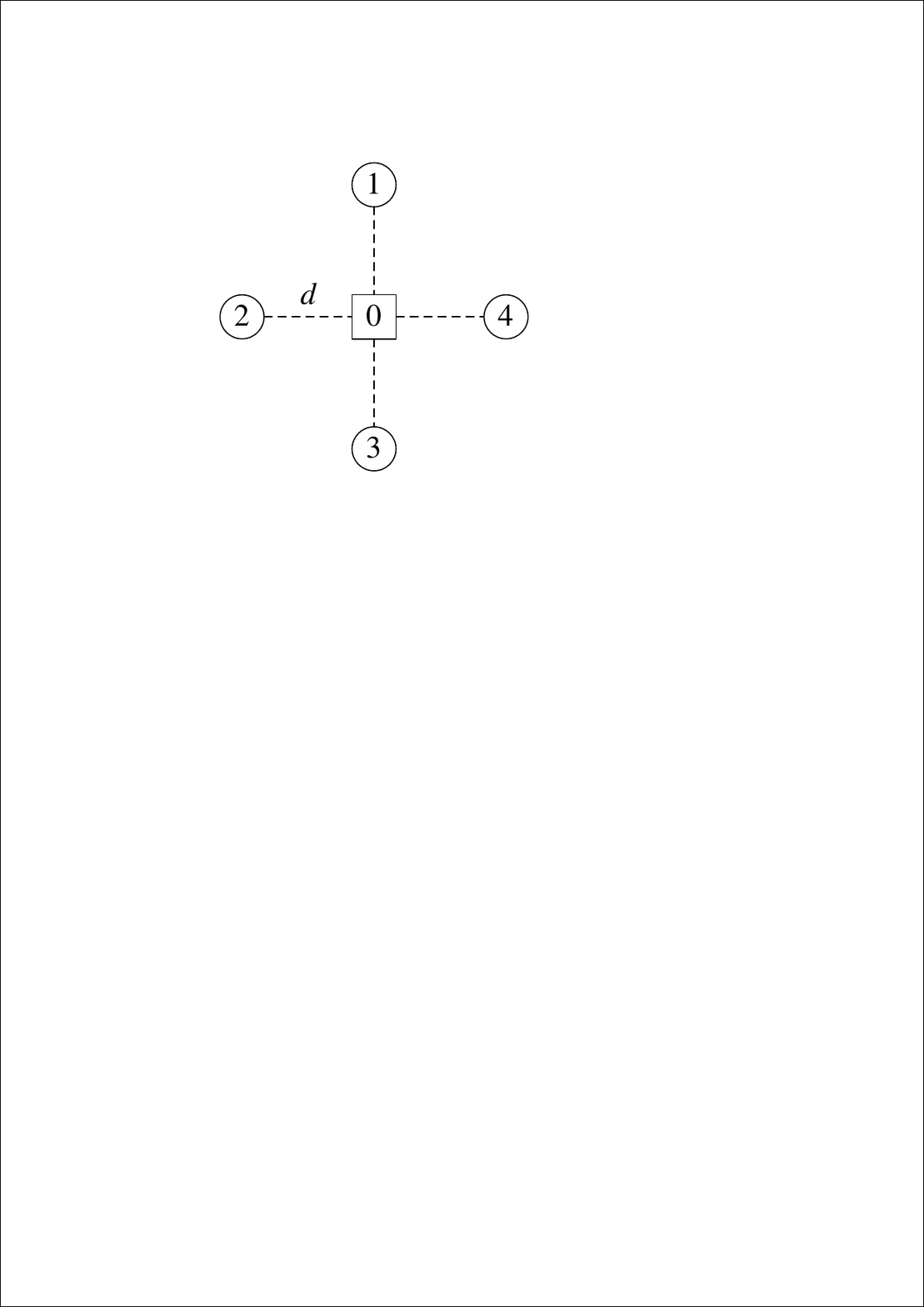}\label{fig:startopology}}
\hfill
\subfigure[Mean of $\TDAoI(1)$]{\includegraphics[width=0.35\linewidth]{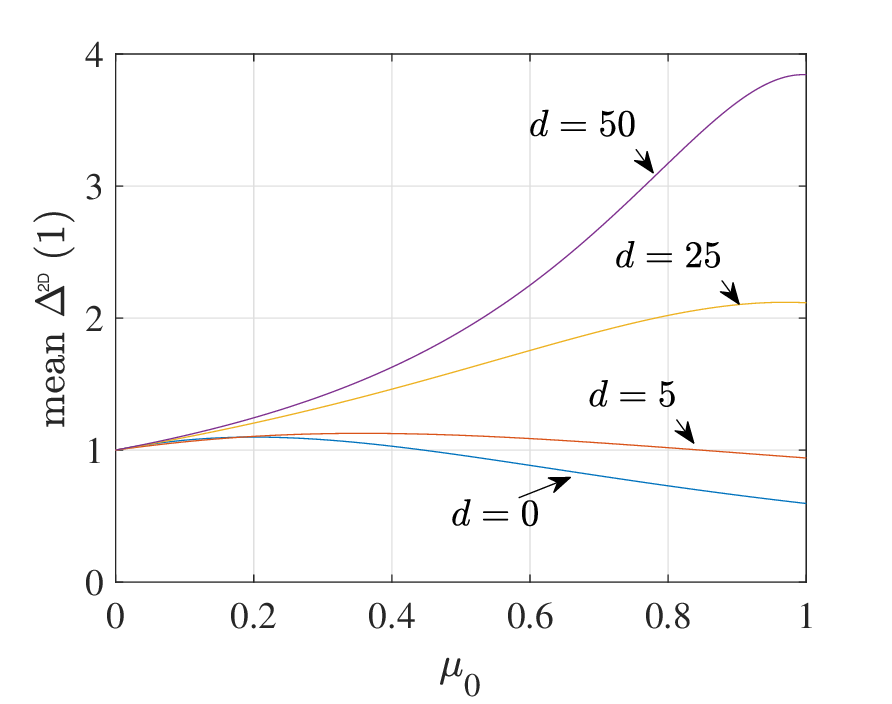}\label{fig:scenario1expected}}
\hfill
\subfigure[0.9 quantile of $\TDAoI(1)$]{\includegraphics[width=0.35\linewidth]{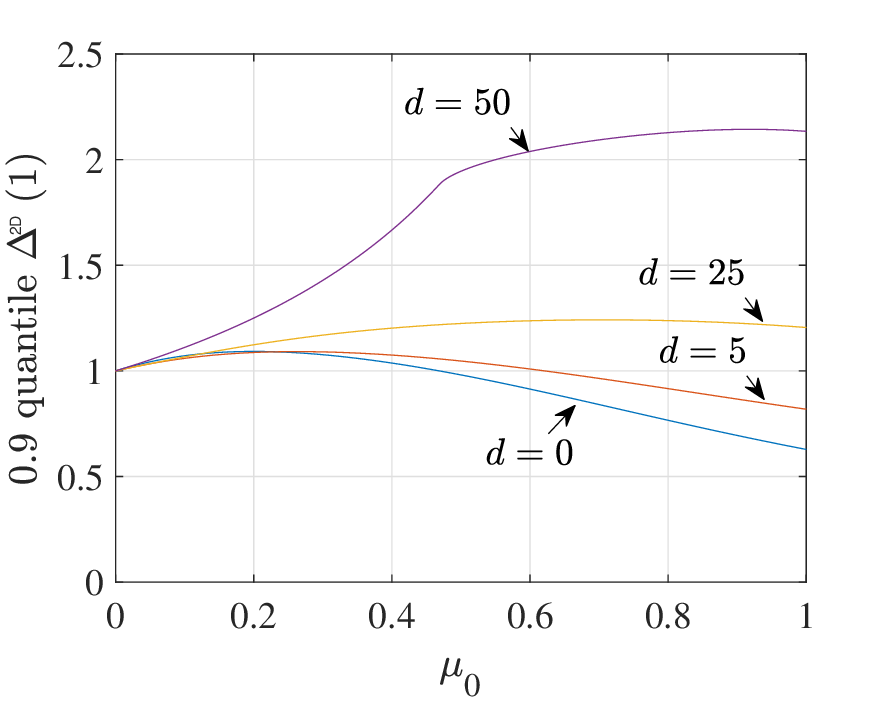}\label{fig:scenario1ccdf90}}
\caption{Minimal \AbbrTDAoI{} $\TDAoI(1)$ for sensor $s=1$ and different spatial distances $d \in \{ 0,5,25,50\}$. To facilitate visual comparison, the results are shown relative to the respective value of the curve at $\mu_0=0$. $\mu_0$ is the service rate allocated to the center node under a sum rate constraint $\sum_s \mu_s = 1$.}
\label{fig:scenario1}
\end{figure*}

In this analysis, we consider four sensors $s \in \{1,2,3,4\}$, which are arranged as a star topology, as illustrated in Fig.~\ref{fig:startopology}. Each of the sensors is connected via a homogeneous, independent M$\mid$M$\mid$1 queue with service rate $\mu_s$. The sensors are positioned at a distance $d$ from the center node. As the star is symmetric, $\TDAoI(s)$ is identical for all $s \in \{1,2,3,4\}$ and it is sufficient to evaluate $\TDAoI(1)$ only. We take the Euclidian distances of the sensors as the AeD that follows as $\Lambda(2,1) = \Lambda(4,1) = \sqrt{2}d$ and $\Lambda(3,1) = 2d$. 

The objective is to find out whether the installation of an additional sensor $s=0$ at the center node is advantageous in this topology, and, if so, how much capacity should be allocated to the center node under a sum rate constraint $\sum_s \mu_s = 1$. Due to homogeneity $\mu_1 = \mu_2 = \mu_3 = \mu_4$ and the sum rate constraint implies $\mu_0 = 1-4\mu_1$. Hence, the question is whether to divide resources and assign them to the leaves $s \in \{1,2,3,4\}$ or pool resources at the center $s=0$. Pooling achieves a faster tail decay of Eq.~\eqref{eq:ageccdfmm1} but at the cost of the AeD of the center node that is $\Lambda(0,1) = d$.

We evaluate the mean and 0.9 quantile of $\TDAoI(1)$ for $\mu_0 \in [0,1]$ in Figs.~\ref{fig:scenario1expected} and~\ref{fig:scenario1ccdf90}, respectively. To simplify the visual representation, we have normalized the results, i.e., we have divided all results by the corresponding value obtained for $\mu_0=0$. For $d=0$ it is evident that pooling is the optimal strategy as $\mu_0=1$ yields the smallest mean $\TDAoI(1)$, see Fig.~\ref{fig:scenario1expected}. Conversely, the least favorable outcome is observed when resources are distributed evenly across all five sensors, i.e., $\mu_0=0.2$. As the distance $d$ increases, the distant sensors including the center node $s \in \{0,2,3,4\}$ become less useful for $\TDAoI(1)$. For $d=5$ pooling, i.e., $\mu_0=1$, remains to be the optimal choice. However, for $d=25$ and $d=50$, it is preferable to divide the resources without using the center node, i.e., $\mu_0 = 0$.

The 0.9 quantile in Fig.~\ref{fig:scenario1ccdf90} is larger in absolute terms but exhibits smaller relative change when $\mu_0$ is varied. For $d=50$ a bend in the 0.9 quantile curve is noticeable at $\mu_0 \approx 0.47$. This phenomenon can be attributed to an effect previously observed in Fig.~\ref{fig:mm1CCDFsAll}. The spatial distance $d=50$ is so large that distant sensors can make a contribution only in rare cases when $\TDAoI(1)$ is very large. Such cases occur with a probability below $10^{-1}$ and are not considered in the 0.9 quantile. This changes when $\mu_0 > 0.47$, as the faster tail decay achieved by the higher service rate can compensate for the spatial distance, resulting in the bend of the curve. Nevertheless, the impact of this effect does not outweigh the disadvantage of the decrease in $\mu_1$ when $\mu_0$ is increased.
%
%
\section{Gaussian Spatio-Temporal Process Model}
\label{sec:processmodel}
We defined \AbbrTDAoI{} $\TDAoI(\varsigma,s)$ in Sec.~\ref{sec:spatiotemporaldistance} as an abstract and generic concept that can be tailored to different applications. In this section, we show a way in which Gaussian processes can be used to instantiate the \AbbrTDAoI{}. A comprehensive treatment of Gaussian process regression is~\cite{rasmussen:gaussianprocesses}. We also give a brief background in Appendix~\ref{appendix:GP}. The use of Gaussian processes also links AoI to Gaussian process regression and machine learning. Depending on the application, alternative process models and metrics may be useful for specifying the \AbbrTDAoI{}. One such case is presented in Appendix~\ref{appendix:random_AeD}.

The physical phenomenon that we consider is a spatio-temporal process that is modeled as a two-dimensional Gaussian process. Additional dimensions, e.g., the correlation of different types of sensors, such as camera, lidar, and radar, can be included as well. We use the posterior variance of the Gaussian process model to establish a link between space and time and to derive the Age-equivalent of the Distance (AeD) of a sensor. The posterior variance is associated with the AoI when the most recent sample of a sensor that is available at the monitor is used for prediction. In the case of spatio-temporal processes (or general higher-dimensional Gaussian processes), we use the posterior variance resulting from the most recent sample of a distant sensor to derive the \AbbrTDAoI{} as the equivalent age (in terms of posterior variance) of a sample at the position of interest.
%
%
\subsection{Prediction Variance}
We characterise the samples that are available at the monitor at time $t \ge 0$ by indicator variables $\Gamma_t(\varsigma,\tau)$, where $\varsigma \in \mathbb{S}$ denotes the sensor and $\tau \in [0,t]$ is the generation time of the sample. Assuming the monitor does not have knowledge of the physical process at position $s$ and time $t$, it uses the samples $\Gamma_t(\varsigma,\tau)$ to make a prediction and we denote $\Phi_t(\Gamma,s)$ as the prediction variance of the estimate. For a Gaussian process we use~\cite[Eq. (2.19)]{rasmussen:gaussianprocesses} to derive the prediction variance 
\begin{equation}
\Phi_t(\Gamma,s) = \Sigma_{{\bf f}_*{\bf f}_*} - \Sigma_{{\bf f}_*{\bf f}} \Sigma^{-1}_{{\bf f}{\bf f}} \Sigma_{{\bf f}{\bf f}_*},
\label{eq:predictionvariance}
\end{equation}
where $\bf{f}$ comprises the samples that are used to make the prediction, these are the samples marked in $\Gamma$, and $\bf{f_*}$ is the prediction, here of the process at position $s$ and time $t$. $\Sigma_{{\bf ff}_*}$ is the $m \times m_*$ covariance matrix of all pairs of values in ${\bf f}$ and ${\bf f}_*$ and so on. Notably, for Gaussian processes, the prediction variance does not depend on the actual values that are observed.

Similarly, we use $\Phi_t(\varsigma,s)$ to denote the prediction variance for position $s$ and time $t$ given only the most recent sample from sensor $\varsigma$ with AoI $\Delta_t(\varsigma)$. The variance $\Phi_t(\varsigma,s)$ can be viewed as a penalty that is due to the spatial distance of the sensor and the AoI of the sample. Conversely, the term $\Sigma_{{\bf f}_*{\bf f}} \Sigma^{-1}_{{\bf f}{\bf f}} \Sigma_{{\bf f}{\bf f}_*}$ is a measure of the value-of-information of the sample, as it quantifies the reduction in uncertainty that is achieved by the sample. Since the definition of AoI is based only on the generation time of the most recent sample available at the monitor, we use the same approach to derive a relation between the posterior variance and the \AbbrTDAoI{}. We note that the monitor can still use all available samples and Eq.~\eqref{eq:predictionvariance} to reduce the prediction variance further.
%
%
\subsection{\AbbrTDAoI{} of Product Kernels}
\label{sec:hdaoi}
Eq.~\eqref{eq:predictionvariance} provides a means of calculating the prediction variance for any given covariance function $k({\bf x},{\bf x}')$, also referred to as the kernel. Evaluating the kernel at defined points in time and space yields the covariance matrices $\Sigma_{{\bf f}{\bf f}_*}$ etc., which are used in Eq.~\eqref{eq:predictionvariance}. In the following, we employ a number of well-established kernel functions in order to derive closed-form results. In particular, we formalize the relationship between space and time, thereby obtaining the AeD of sensors. We use stationary product kernels~\cite{Duvenaud:kernelcookbook} to model spatio-temporal processes. To include a range of common kernel functions, we consider kernels with the following structure
\begin{equation}
k(\varsigma,s,\Delta_t) = \sigma^2 g(\Delta_t(\varsigma)) h(\varsigma,s) ,
\label{eq:generalexpokernelfunctions}
\end{equation}
where functions $g$ and $h$ satisfy $g(0) = h(s,s) = 1$ for all $s \in \mathbb{S}$, and $\sigma^2$ is the process variance. Function $g$ denotes the temporal correlation and $h$ is the spatial correlation. Additional dimensions, such as non-homogeneous sensors, can be included in $h$ by extending the index set, or represented by multiplication of additional kernel functions. Well-known examples of Eq.~\eqref{eq:generalexpokernelfunctions} are the exponential, the squared exponential, and the rational quadratic kernels.

In line with the general assumption of AoI, that a new sample replaces previous samples, we assume that function $g$ is decreasing. This means that a more recent sample is better suited for prediction than any previous one. We define $g^{-1}$ to be the inverse of $g$. Since $g$ is decreasing, it follows that $g^{-1}$ is decreasing, too. To see this, define $y_i=g(x_i)$ and $x_i=g^{-1}(y_i)$. Then for $g$ decreasing and $y_1 < y_2$
\begin{equation*}
y_1 < y_2 \Leftrightarrow g(x_1) < g(x_2) \Leftrightarrow x_1 > x_2 \Leftrightarrow g^{-1}(y_1) > g^{-1}(y_2) .
\end{equation*}

The prediction variance given the most recent sample of sensor $\varsigma$ follows by insertion of Eq.~\eqref{eq:generalexpokernelfunctions} into Eq.~\eqref{eq:predictionvariance} as 
\begin{equation}
\Phi_t(\varsigma,s) = \sigma^2 \Bigl( 1- \bigl(g(\Delta_t(\varsigma))h(\varsigma,s)\bigl)^2 \Bigr),
\label{eq:predictionvariancesinglesample}
\end{equation}
where we used that $\Sigma_{\bf ff} = \Sigma_{\bf f_*f_*} = \sigma^2$, $\Sigma_{\bf ff}^{-1} = 1/\sigma^2$, and $\Sigma_{\bf ff_*} = \Sigma_{\bf f_*f} = \sigma^2  g(\Delta_t(\varsigma))h(\varsigma,s)$.

The next step is to leverage the structure of the prediction variance in Eq.~\eqref{eq:predictionvariancesinglesample} to develop a definition of \AbbrTDAoI{} that transforms space into an equivalent of time. This is achieved by setting
\begin{equation}
\TDAoI_t(\varsigma,s) := g^{-1}( g(\Delta_t(\varsigma))h(\varsigma,s) ) .
\label{eq:spatiotemporaldistancekernel}
\end{equation}
The definition implies that 
\begin{equation*}
g(\TDAoI_t(\varsigma,s)) = g(\Delta_t(\varsigma))h(\varsigma,s) ,
\end{equation*}
i.e., the \AbbrTDAoI{} transfers the spatial correlation $h(\varsigma,s)$ to the temporal domain, expressed by function $g(\TDAoI_t(\varsigma,s))$. Thus, Eq.~\eqref{eq:predictionvariancesinglesample} can be written as
\begin{equation}
\Phi_t(\varsigma,s) = \sigma^2 \Bigl( 1- \bigl(g(\TDAoI_t(\varsigma,s))\bigr)^2\Bigr),
\label{eq:predictionvariancesinglesamplesolved}
\end{equation}
using only the temporal correlation function $g$. This means that for predicting the physical phenomenon at position $s$ and time $t$, a sample from position $\varsigma$ with AoI $\Delta_t(\varsigma)$ is as good as using a sample from position $s$ with AoI $\Delta_t(s) = \TDAoI_t(\varsigma,s)$. This achieves the intended transformation of the spatial distance into an age-equivalent representation. Given the prediction variance, Eq.~\eqref{eq:predictionvariancesinglesamplesolved} can be solved for the \AbbrTDAoI{}
\begin{equation*}
\TDAoI_t(\varsigma,s) = g^{-1}\left(\sqrt{1- \frac{\Phi_t(\varsigma,s)}{\sigma^2}} \right) .
\end{equation*}

As defined in Eq.~\eqref{eq:spatialdistance} we have $\TDAoI_t(\varsigma,s) = \Lambda_t(\varsigma,s) + \Delta_t(\varsigma)$ and by insertion of Eq.~\eqref{eq:spatiotemporaldistancekernel} into Eq.~\eqref{eq:spatialdistance} we obtain the AeD
\begin{equation}
\Lambda_t(\varsigma,s) = g^{-1}( g(\Delta_t(\varsigma))h(\varsigma,s) ) - \Delta_t(\varsigma) .
\label{eq:lambda}
\end{equation}
We will derive $\Lambda_t(\varsigma,s)$ and hence $\TDAoI_t(\varsigma,s)$ for different commonly used kernel functions in Sec.~\ref{sec:catalogeofkernels}, where we find that in some but not all cases $\Lambda_t(\varsigma,s)$ is a function of space only, i.e., $\Lambda_t(\varsigma,s) = \Lambda(\varsigma,s)$ as depicted in Fig.~\ref{fig:spatiotemporaldistance}. 

Before we move on to specific kernel functions, we conclude this section with some general results. With Eq.~\eqref{eq:spatialdistance} the CCDF of $\TDAoI_t(\varsigma,s)$ is
\begin{equation}
\mathsf{P} [\TDAoI_t(\varsigma,s) > y] = \mathsf{P} [\Delta_t(\varsigma) > y - \Lambda_t(\varsigma,s)] ,     
\label{eq:spatiotemporaldistanceccdf}
\end{equation}
for $y \ge \Lambda_t(\varsigma,s)$. The CCDF of the prediction variance follows from Eq.~\eqref{eq:predictionvariancesinglesamplesolved} as
\begin{align*}
\mathsf{P} [ \Phi_t(\varsigma,s) > z \sigma^2 ]  &= \mathsf{P} [ 1-g(\TDAoI_t(\varsigma,s))^2 > z] \\
&= \mathsf{P} \bigl[ g(\TDAoI_t(\varsigma,s)) < \sqrt{1-z}\bigr] \nonumber \\
&= \mathsf{P} \bigl[ \TDAoI_t(\varsigma,s) > g^{-1} \bigl(\sqrt{1-z}\bigr)\bigr] ,
\end{align*}
where we normalized the CCDF with the variance, so that $z \in [0,1]$. Above, since $g^{-1}$ is a decreasing function the order of the inequality is inverted. Solving for $\Delta_t(\varsigma)$ the CCDF of the prediction variance can also be written as
\begin{align}
\mathsf{P} [ \Phi_t(\varsigma,s) > z \sigma^2 ] &= \mathsf{P} \biggl[ \Delta_t(\varsigma) > g^{-1} \biggl(\frac{\sqrt{1-z}}{h(\varsigma,s)}\biggr)\biggr] .
\label{eq:predictionvarianceccdf}
\end{align}
%
%
\subsection{Observation Noise}
\label{sec:observationnoise}
Assuming noisy observations {\bf f} with additive iid Gaussian noise $\epsilon$, where $\epsilon$ has zero mean and variance $\sigma_N^2$, the covariance of the observations becomes $\Sigma_{\bf f \bf f} = k({\bf x}, {\bf x'}) + I \sigma_N^2$, see e.g.~\cite[Eq. (2.20)]{rasmussen:gaussianprocesses}. The prediction variance follows by insertion of $\Sigma_{\bf f \bf f}$ into Eq.~\eqref{eq:predictionvariance}. Here we assume that the noise of all sensors is homogeneous with variance $\sigma_N^2$. Sensors with different noise variance can be taken into account at the expense of additional notation. When using the most recent sample of sensor $\varsigma$ for prediction, we have $\Sigma_{\bf f \bf f} = \sigma^2 + \sigma_N^2$ and by insertion of Eq.~\eqref{eq:generalexpokernelfunctions} into Eq.~\eqref{eq:predictionvariance} the prediction variance turns into
\begin{equation}
\Phi_t(\varsigma,s) = \sigma^2 \bigl(1 - \eta  ( g(\Delta_t(\varsigma))h(\varsigma,s) )^2\bigr),
\label{eq:predictionvariancesinglesamplenoise}
\end{equation}
with $\eta = \frac{\sigma^2}{\sigma^2 + \sigma_N^2}$. Here we used that $\Sigma_{\bf f_*f_*} = \sigma^2$, $\Sigma_{\bf ff}^{-1} = 1/(\sigma^2 + \sigma_N^2)$, and $\Sigma_{\bf ff_*} = \Sigma_{\bf f_*f} = \sigma^2 g(\Delta_t(\varsigma))h(\varsigma,s)$.

Substituting Eq.~\eqref{eq:predictionvariancesinglesamplenoise} into the key steps of the definition of \AbbrTDAoI{} in Eqs.~\eqref{eq:predictionvariancesinglesample}-\eqref{eq:predictionvariancesinglesamplesolved}, we see that the inclusion of observation noise has no effect on the definition of \AbbrTDAoI{}, but only on the prediction variance, and Eq.~\eqref{eq:predictionvarianceccdf} is shifted so that
\begin{equation*}
\mathsf{P}[\Phi_t(\varsigma,s) > z \sigma^2 ] = \mathsf{P} \biggl[\Delta_t(\varsigma) > g^{-1} \biggl(\frac{\sqrt{1-z}}{\sqrt{\eta} h(\varsigma,s)} \biggr) \biggr] .  
\end{equation*} 
This allows our model to easily incorporate independent Gaussian noise. For convenience, we assume noise-free observations in the following, i.e. we use $\eta = 1$.
%
%
\section{\AbbrTDAoI{} of Common Kernels}
\label{sec:catalogeofkernels}
Using the general results from Sec.~\ref{sec:processmodel} we derive specific solutions for products of some widely used kernel functions including exponential, squared exponential, and rational quadratic. Empirical data have often been used to parameterize these kernels, e.g., exponential product kernels have been used to model spatio-temporal temperature and humidity data in~\cite{hribar2018using}, squared exponential kernels for road speed correlation~\cite{krause:communitysensing}, and product kernels including squared exponential and rational quadratic for atmospheric CO$_2$ concentration~\cite{rasmussen:gaussianprocesses}, to name a few. We discuss the cases of rational quadratic kernels and mixed product kernels in Appendix~\ref{appendix:rational_quadratic_kern} and~\ref{appendix:mixed_product_kern} to demonstrate the broader applicability of our method.
%
%
\subsection{Exponential Kernels}
For the case of exponential product kernels we instantiate Eq.~\eqref{eq:generalexpokernelfunctions} with $g(\Delta_t(\varsigma)) = e^{-\frac{1}{l_t} \Delta_t(\varsigma)}$ and $h(\varsigma,s) = e^{-\frac{1}{l_s} |x_{\varsigma}-x_s|}$, where $l_s, l_t > 0$ are the length scales of the spatial and temporal correlation, respectively. The inverse function of the temporal correlation function follows as
\begin{equation*}
g^{-1}(y) = -l_t \ln y ,
\end{equation*}
for $y \in (0,1]$, and by insertion into Eq.~\eqref{eq:lambda} the AeD is
\begin{align}
\Lambda (\varsigma,s) &=  -l_t \ln \Bigl( e^{-\frac{1}{l_s} |x_{\varsigma}-x_s|}  e^{-\frac{1}{l_t} \Delta_t(\varsigma)} \Bigr) - \Delta_t(\varsigma) \nonumber \\
&= \frac{l_t}{l_s} |x_{\varsigma}-x_s| .
\label{eq:spatialdistanceexpo}
\end{align}
The CCDF of the \AbbrTDAoI{} follows directly by insertion of Eq.~\eqref{eq:spatialdistanceexpo} into Eq.~\eqref{eq:spatiotemporaldistanceccdf} as
\begin{equation}
\mathsf{P}[\TDAoI_t (\varsigma,s) > y] = \mathsf{P}\biggl[\Delta_t(\varsigma) > y - \frac{l_t}{l_s} |x_{\varsigma}-x_s|\biggr]  ,
\label{eq:hdaoiexpo}
\end{equation}
for $y \ge \frac{l_t}{l_s} |x_{\varsigma}-x_s|$. The CCDF of the prediction variance from Eq.~\eqref{eq:predictionvarianceccdf} is
\begin{align}
\mathsf{P} [ \Phi_t(\varsigma,s) > z \sigma^2 ] &= \mathsf{P} \biggl[ \Delta_t(\varsigma) > -l_t \ln \biggl(\frac{\sqrt{1-z}}{ e^{-\frac{1}{l_s} |x_{\varsigma}-x_s|} }\biggr)\biggr] \nonumber \\
&= \mathsf{P} \biggl[ \Delta_t(\varsigma) > - \frac{l_t}{2} \ln(1-z) - \frac{l_t}{l_s} |x_{\varsigma}-x_s| \biggr]  ,
\label{eq:predvarexpo}
\end{align}
for $z \ge 1-e^{-\frac{2}{l_s} |x_{\varsigma}-x_s|}$ and $z < 1$.
For exponential product kernels, the AeD $\Lambda(\varsigma,s)$ in Eq.~\eqref{eq:spatialdistanceexpo} is a function of space only. This is the basic case that we depicted already in Fig.~\ref{fig:spatiotemporaldistance}. Numerical results for this case have been presented in Sec.~\ref{sec:numericalexample1} and Sec.~\ref{sec:numericalexample2}. 
%
%
\subsection{Squared Exponential Kernels}
\label{sec:catalogeofkernels_SE}
\begin{figure}
\subfigure[Sample path of the \AbbrTDAoI{}]{\includegraphics[width=0.5\linewidth]{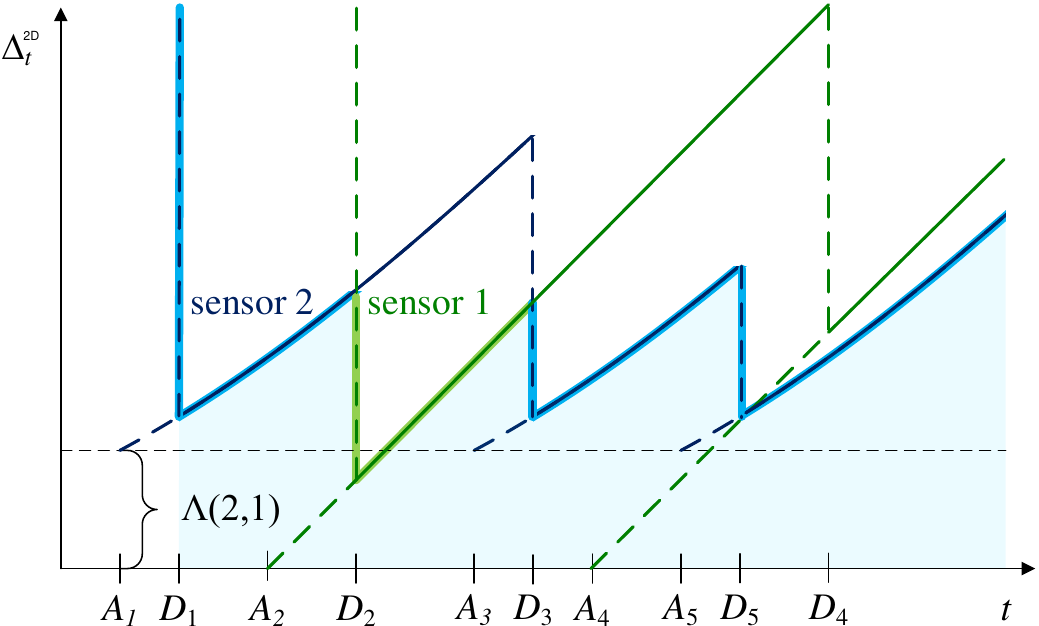}\label{fig:spatiotemporaldistancequadratic}}
\hfill
\subfigure[CCDF of the \AbbrTDAoI{}]{\includegraphics[width=0.41\linewidth]{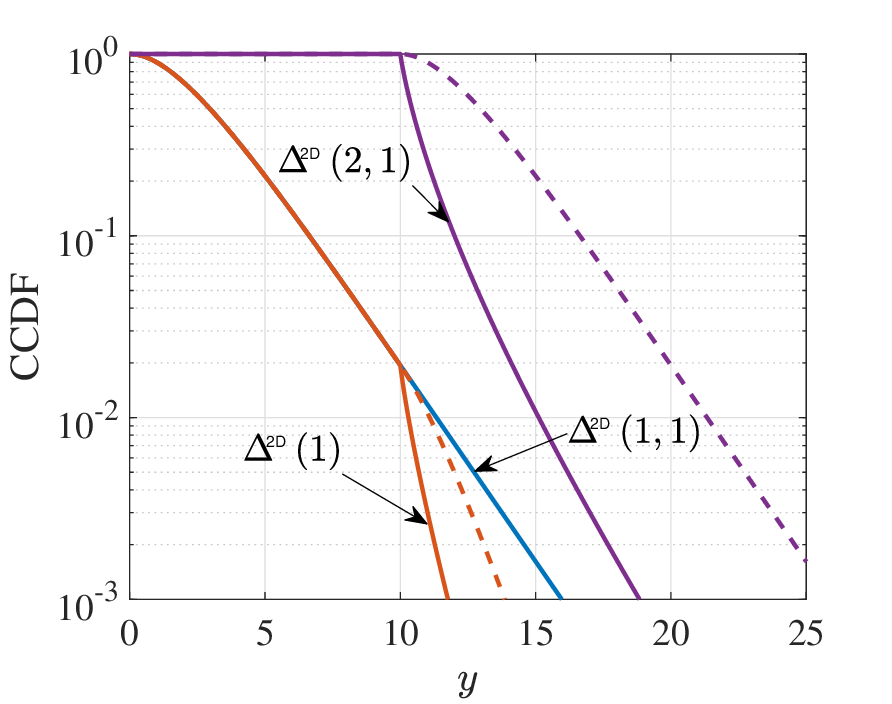}\label{fig:mm1ccdfs_se}}
\caption{Fig.~\ref{fig:spatiotemporaldistancequadratic} shows the minimal \AbbrTDAoI{} $\TDAoI_t(1)$ for the case of $S=2$ sensors as in Fig.~\ref{fig:spatiotemporaldistance} but for squared exponential instead of exponential kernels. Fig.~\ref{fig:mm1ccdfs_se} shows the CCDF like Fig.~\ref{fig:mm1CCDFsAll} but with squared exponential kernels. The dashed lines show the corresponding result when using an exponential kernel.}
\label{fig:quadratickernel}
\end{figure}

For squared exponential kernels we substitute $g(\Delta_t(\varsigma)) = e^{-\frac{1}{2l_t^2} (\Delta_t(\varsigma))^2}$ and $h(\varsigma,s) = e^{-\frac{1}{2l_s^2} (x_{\varsigma}-x_s)^2}$ into Eq.~\eqref{eq:generalexpokernelfunctions}. It follows that 
\begin{equation*}
g^{-1}(y) = l_t \sqrt{-2 \ln y} ,
\end{equation*}
for $y \in (0,1]$. By insertion into Eq.~\eqref{eq:lambda} we obtain
\begin{align}
\Lambda_t(\varsigma,s) &= \biggl( -2l_t^2 \ln \biggl( e^{-\frac{1}{2l_s^2} (x_{\varsigma}-x_s)^2}  e^{-\frac{1}{2l_t^2} \Delta_t(\varsigma)^2} \biggr)\biggr)^{\frac{1}{2}} - \Delta_t(\varsigma) \nonumber \\
&= \sqrt{\frac{l_t^2}{l_s^2} (x_{\varsigma}-x_s)^2  + \Delta_t(\varsigma)^2} - \Delta_t(\varsigma) .
\label{eq:spatialdistancesquaredexpo}
\end{align}
In comparison to the behavior observed in Eq.~\eqref{eq:spatialdistanceexpo}, the behavior exhibited by Eq.~\eqref{eq:spatialdistancesquaredexpo} is quite distinct. In this case, the AeD $\Lambda_t(\varsigma,s)$ is a function not only of space but also of time, specifically the AoI at the given time point $t$. When $\Delta_t(\varsigma) = 0$, Eq.~\eqref{eq:spatialdistancesquaredexpo} is identical to Eq.~\eqref{eq:spatialdistanceexpo}. However, the magnitude of $\Lambda_t(\varsigma,s)$ in Eq.~\eqref{eq:spatialdistancesquaredexpo} is reduced as $\Delta_t(\varsigma)$ increases, with $\Lambda_t(\varsigma,s) \to 0$ for $\Delta_t(\varsigma) \to \infty$. The effect is illustrated in the schematic in Fig.~\ref{fig:spatiotemporaldistancequadratic}, which can be directly compared with Fig.~\ref{fig:spatiotemporaldistance}. In contrast to Fig.~\ref{fig:spatiotemporaldistance}, the spatial offset of the blue curve representing sensor 2 gradually diminishes over time in Fig.~\ref{fig:spatiotemporaldistancequadratic}. This indicates that when sensors communicate less frequently, samples from distant sensors are of greater value in the case of squared exponential kernels.

The CCDF of the \AbbrTDAoI{} follows by insertion of Eq.~\eqref{eq:spatialdistancesquaredexpo} into Eq.~\eqref{eq:spatiotemporaldistanceccdf} as
\begin{equation}
\mathsf{P}[\TDAoI_t(\varsigma,s) > y] = \mathsf{P} \Biggl[\Delta_t(\varsigma) > \biggl(y^2 - \frac{l_t^2}{l_s^2} (x_{\varsigma}-x_s)^2\biggr)^{\frac{1}{2}} \Biggr] ,
\label{eq:hdaoisquaredexpo}
\end{equation}
for $y \ge \frac{l_t}{l_s} |x_\varsigma-x_s|$, and the CCDF of the minimal \AbbrTDAoI{} is obtained by insertion into Eq.~\eqref{eq:spatiotemporaldistanceccdfgeneral}. 
The CCDF of the prediction variance in Eq.~\eqref{eq:predictionvarianceccdf} of the squared exponential kernel becomes
\begin{align}
\mathsf{P} [ \Phi_t(\varsigma,s) > z \sigma^2 ] &= \mathsf{P} \left[ \Delta_t(\varsigma) > \Biggl(- 2l_t^2 \ln \Biggl(\frac{\sqrt{1-z}}{ e^{-\frac{1}{2l_s^2} (x_{\varsigma}-x_s)^2} }\Biggr)\Biggr)^{\frac{1}{2}}\right] \nonumber \\
&= \mathsf{P} \left[ \Delta_t(\varsigma) > \biggl(- l_t^2 \ln(1-z)  - \frac{l_t^2}{l_s^2} (x_{\varsigma} - x_s)^2 \biggr)^{\frac{1}{2}} \right],
\label{eq:predvarsquaredexpo}
\end{align}
for $z \ge 1-e^{-\frac{1}{l_s^2} (x_{\varsigma}-x_s)^2}$ and $z < 1$.

Results for the scenario evaluated in Sec.~\ref{sec:numericalexample1}, where two sensors are spaced apart and connected to the monitor via independent M$\mid$M$\mid$1 queues, are shown in Fig.~\ref{fig:mm1ccdfs_se} for squared exponential kernels. The parameters of the kernels are selected so that $(x_2-x_1)^2 l_t^2/l_s^2 = 100$, i.e., for $\Delta_t(2) = 0$ Eq.~\eqref{eq:spatialdistancesquaredexpo} gives an AeD of $\Lambda_t(2,1) = 10$. For comparison, the dashed lines show the previous results from Fig.~\ref{fig:mm1CCDFsAll} that apply for exponential kernels with the same parameters $x_1,x_2,l_t,l_s$. 

Fig.~\ref{fig:mm1ccdfs_se} clearly demonstrates the impact of the squared exponential kernel compared to the exponential kernel. The offset of $\TDAoI(2,1)$ for $\Delta_t(2) = 0$ is identical to $y=10$ in both cases. However, for $y > 10$, that is when $\Delta_t(2)$ increases, the CCDF of $\TDAoI(2,1)$ of the squared exponential kernel decays more rapidly. This is a consequence of Eq.~\eqref{eq:spatialdistancesquaredexpo}. As the value of $y$ increases, the CCDFs of $\TDAoI(2,1)$ and $\TDAoI(1,1)$ converge for the squared exponential kernel, showing the decreasing effect of the distance of sensor 2. The CCDF of $\TDAoI(2,1)$ of the exponential kernel (dashed lines) has the same speed of tail decay but will remain separated from $\TDAoI(1,1)$ by $y=10$. While $\TDAoI(1,1)$ is identical for both types of kernels, since it depends on the AoI $\Delta(1)$ only, the minimal \AbbrTDAoI{} $\TDAoI(1)$ benefits from the more rapid decay of $\TDAoI(2,1)$ for the squared exponential kernel, beginning at $y=10$ and a tail probability of about 0.02.
%
%
\section{Evaluation of the Spatial Density of Sensors}
\label{sec:evaluation}
In this section, we study the \AbbrTDAoI{} of a sensor grid that monitors a physical phenomenon modeled as a spatio-temporal Gaussian process. The sensor nodes are connected to a monitor via independent channels that are modeled as M$\mid$M$\mid$1 queues in Sec.~\ref{sec:evaluationmm1} and as slotted ALOHA channels in Sec.~\ref{sec:evaluationaloha}, where we also consider non-independent channels. Other channel models are also possible, only the CCDF of the AoI of that model is required. We also include results of the prediction variance in Sec.~\ref{sec:evaluationpredictionvariance}.
In general, we assume a per-area capacity constraint which means that if the spatial density of the sensors is increased, the transmission capacity per sensor decreases, thereby reducing the temporal intensity of sampling. We evaluate this trade-off between spatial and temporal sampling intensity and present results showing the optimal density of sensors. 

\begin{figure}
\centering
\includegraphics[width=0.25\linewidth]{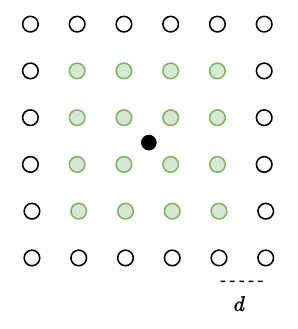}
\caption{Regular sensor grid. An area with $N=36$ sensors (white and green dots) is shown. The sensors have a sum rate constraint determined by the per-area capacity. The distance between the sensors is $d$. A smaller $d$ leads to a larger $N$, reducing the capacity per sensor. The monitor predicts the process at the point of interest (black dot) using samples of the nearest $S$ sensors, here $S=16$ (green dots).}
\label{fig:grid36_2}
\end{figure}
In detail, we examine a regular two-dimensional sensor grid, as illustrated in Fig.~\ref{fig:grid36_2}, where the distance between neighboring sensors is denoted $d$. Random sensor topologies, such as the one-dimensional Poisson point process studied in~\cite{jiang2019howdensely}, can be taken into account by treating the AeD $\Lambda_t(\varsigma,s)$ as a random variable (due to random sensor positions $x_{\varsigma}$). To avoid boundary effects, we consider an infinite grid of sensors. The sensor grid is partitioned into square areas of size $A$ and the sensor distance $d$ determines the number of sensors $N$ per area, denoted by white and green dots in Fig.~\ref{fig:grid36_2}. In the evaluation, we set $A=300^2$, which is a typical size of a wireless local area network when the units are meters. The $N$ sensors within each area are subject to a sum rate constraint, which is determined by the per-area capacity. The areas are statistically equivalent, and thus it is sufficient to consider a single area. 

We select the center of the area, which is marked by a black dot in Fig.~\ref{fig:grid36_2}, as the point of interest. This point maximizes the distance to the nearest sensor. To predict the process at the point of interest, the samples of the $S \le N$ nearest sensors in the area are used by the monitor. We evaluated different values of $S$ and the results shown here are for $S=16$ sensors marked by green dots in Fig.~\ref{fig:grid36_2}. We refer to the inner 4 sensors next to the point of interest as tier-1 and to the 12 sensors surrounding the tier-1 sensors as tier-2 and so on. Generally, increasing $S$ will improve the \AbbrTDAoI{}. However, the effect of larger $S$, including tier-3 sensors and higher, is diminishing in most of our experiments. The reason for this is the greater distance of these sensors to the point of interest, which is reflected in a larger AeD. Tier-3 sensors can only make a useful contribution if the \AbbrTDAoI{} of all $S=16$ tier-1 and tier-2 sensors exceeds the AeD of the tier-3 sensors, and so on for higher tiers. This relationship of \AbbrTDAoI{} and AeD is also shown in Fig.~\ref{fig:mm1CCDFs} for the case of two sensors and we present further results for different $S$ that break down the contribution of the different sensor tiers in Appendix~\ref{appendix:sensor_tiers}. In the evaluation we vary the sensor distance $d$. The distances of the nearest $S=16$ sensors to the point of interest are $d/\sqrt{2}$ for the tier-1 sensors, and $\sqrt{5} d/\sqrt{2}$, respectively, $3 d/\sqrt{2}$ for the tier-2 sensors.

We evaluate two models of the spatio-temporal physical phenomenon, a Gaussian process with an exponential and one with a squared exponential product kernel. The temporal correlation is fixed to a value of $l_t=128$ while we consider different values for the spatial correlation $l_s \in \{64,128,256,512,\infty\}$. The variance of the process is set to $\sigma^2=1$.
%
%
\subsection{Sensors Connected via M$\mid$M$\mid$1 Queues}
\label{sec:evaluationmm1}
\begin{figure*}
\subfigure[$\mu=10$]{\includegraphics[width=0.32\linewidth]{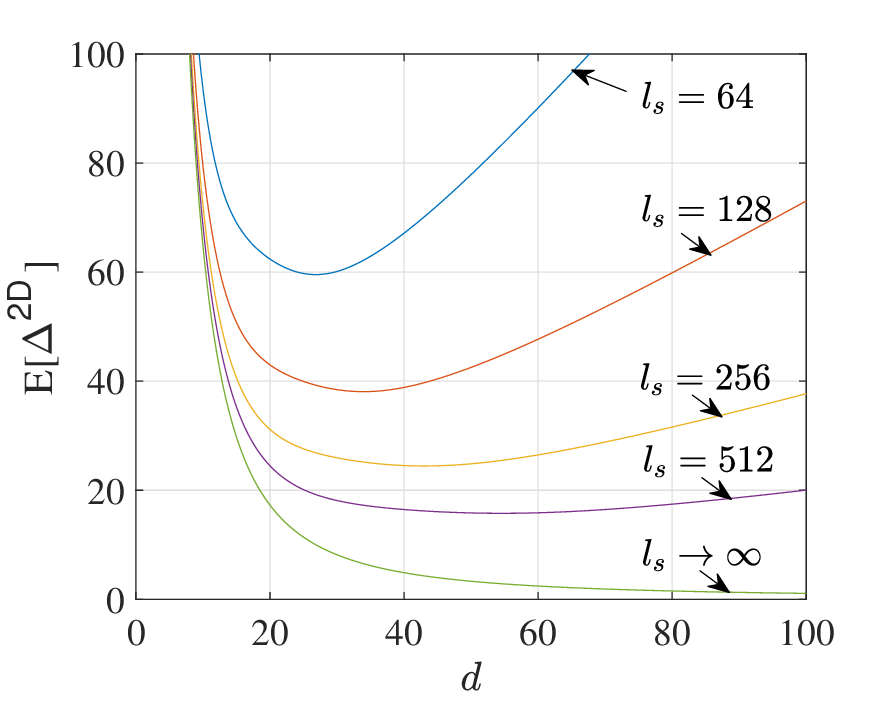}\label{fig:evaluation_AOIvsDistanceMM1mu10}}
\hfill
\subfigure[$\mu=100$]{\includegraphics[width=0.32\linewidth]{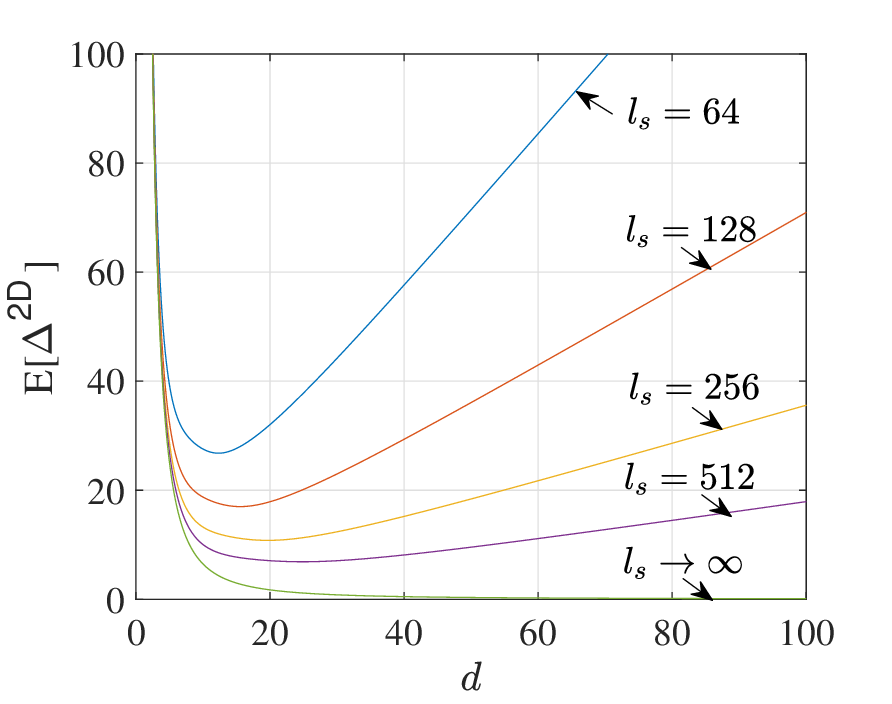}\label{fig:evaluation_AOIvsDistanceMM1mu100}}
\hfill
\subfigure[$\mu=1000$]{\includegraphics[width=0.32\linewidth]{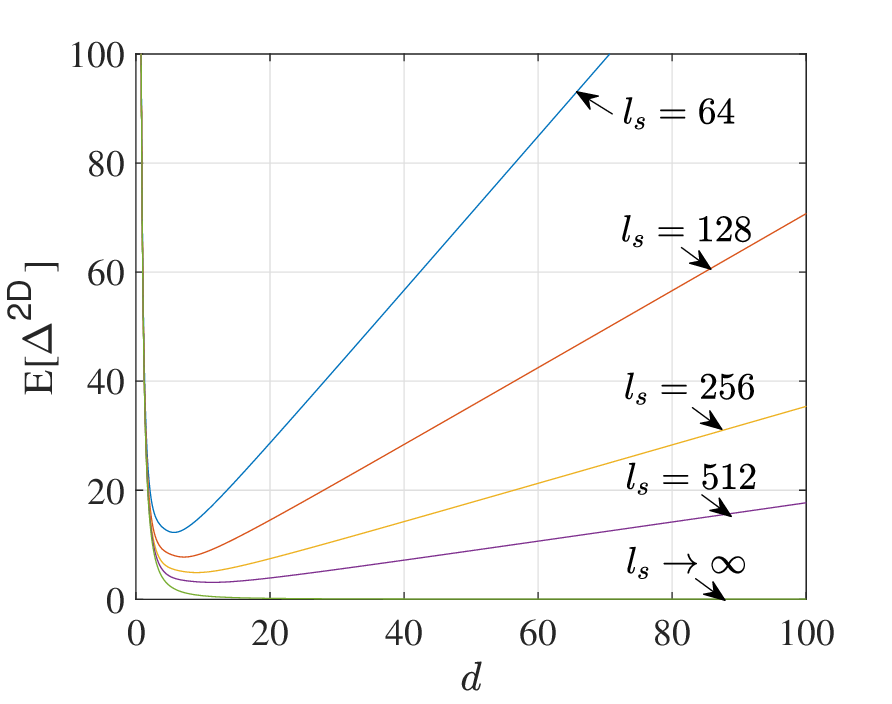}\label{fig:evaluation_AOIvsDistanceMM1mu1000}}
\caption{Monitoring of a spatio-temporal Gaussian process with exponential kernel, temporal correlation coefficient $l_t=128$ and different spatial correlation coefficients $l_s$. Sensors are placed at a distance $d$ and are connected via independent M$\mid$M$\mid$1 queues for different per-area service rates $\mu$ and an area $A=300^2$.}
\label{fig:evaluation_AOIvsDistanceMM1}
\end{figure*}

First, we consider an exponential product kernel and evaluate the case where each sensor sends its samples via an independent M$\mid$M$\mid$1 queue. There is an overall service rate of $\mu$ per area $A$ such that each of the sensors $i \in \{1,2,...,N\}$ in the area has access to an independent queue with service rate $\mu_i=\mu/N$. The sampling rate of each sensor equals the arrival rate to the queue that is $\lambda_i = \rho \mu_i$, where $\rho$ is the utilization. We set $\rho=0.53$, which is the value that minimizes the AoI of the M$\mid$M$\mid$1 queue. We consider distance values in the range of $d \in \{1,...,100\}$ and per-area capacity values $\mu \in \{10,100,1000\}$. Numerical results are provided by substituting Eq.~\eqref{eq:ageccdfmm1} for the M$\mid$M$\mid$1 queue into Eq.~\eqref{eq:hdaoiexpo} for the exponential kernel. The expected value is obtained by integration of the tail~\eqref{eq:meanspatiotemporalpenalty}. The results are shown in Fig.~\ref{fig:evaluation_AOIvsDistanceMM1}.

In all cases in Fig.~\ref{fig:evaluation_AOIvsDistanceMM1}, a significant influence of the sensor distance $d$ on the expected \AbbrTDAoI{} $\mathsf{E}[\TDAoI]$ can be seen, which has a characteristic u-shape. The \AbbrTDAoI{} is composed of AoI and AeD, and the u-shape of the \AbbrTDAoI{} is a consequence of the fact that the AoI is a decreasing and the AeD an increasing function of $d$. If $d$ is small, the spatial density of sensors is high and there are sensors in the vicinity of the point of interest such that the AeD is small. However, due to the high density of sensors and the fixed capacity per area, the service rate per sensor is low, as is the sampling rate. As illustrated by Eq.~\eqref{eq:ageccdfmm1}, a small $\mu_i$ as well as a small $\lambda_i$ result in a considerable increase of the AoI. In comparison, the AeD is insignificant so that the AoI dominates the \AbbrTDAoI{} for small $d$.

In contrast, when $d$ is large, the spatial density of sensors is low, yet the sampling intensity is high, as more capacity is available to each sensor. As a result, the impact of the AoI decreases significantly, while the AeD becomes considerable and dominates the \AbbrTDAoI{}. The AeD is obtained from Eq.~\eqref{eq:spatialdistanceexpo} as $\Lambda = |x_{\varsigma}-x_s| (l_t/l_s)$ and for the sensors in closest proximity to the point of interest, i.e. the tier-1 sensors, $\Lambda = (d/\sqrt{2}) (l_t/l_s)$. At the greatest sensor distance $d=100$, the AoI is very small and the \AbbrTDAoI{} curves converge towards the AeD of the tier-1 sensors that is $\Lambda \approx 71 (128/l_s)$, respectively, $71$ for $l_s = 128$. Due to their larger AeD (by a factor of $\sqrt{5}$ or 3), the tier-2 sensors cannot make a noticeable contribution in this case. The convergence towards the AeD is faster when $\mu$ is larger, see Fig.~\ref{fig:evaluation_AOIvsDistanceMM1mu1000}, but it is also clearly visible in Figs.~\ref{fig:evaluation_AOIvsDistanceMM1mu100} and~\ref{fig:evaluation_AOIvsDistanceMM1mu10}, where $\mu$ is smaller. For small $\mu$ the influence of the AoI remains, however, more noticeable. 

Between the two extremes of very small and very large sensor distances, the ratio of AoI and AeD is more balanced and we can determine a sensor distance $d$ that minimizes the mean \AbbrTDAoI{} for the given parameters (except for the special case $l_s \to \infty$, in which the spatial distance is irrelevant). In a concrete example for $\mu=10$ and $l_s=128$ in Fig.~\ref{fig:evaluation_AOIvsDistanceMM1mu10}, the minimal expected \AbbrTDAoI{} of $38$ is found at a sensor distance of $d=34$. Here, the AeD is $24$ for the tier-1 sensors, and $53.8$ and $72.1$, respectively, for the tier-2 sensors. The AeD of the tier-2 sensors is significantly larger than the expected \AbbrTDAoI{}. Therefore, they can only affect the tail distribution of the \AbbrTDAoI{} and have limited influence on the expected \AbbrTDAoI{}. The tier-2 sensors start contributing more frequently left of the minimum, when for $d < 25.3$ the AeD of the tier-2 sensors becomes smaller than the expected \AbbrTDAoI{} of $40$. Fig.~\ref{fig:evaluation_AOIvsDistanceMM1mu10_tier1_tier2_tier3} in the appendix breaks down these effects by the tier of the sensors.

It is evident that an increase in the area capacity $\mu$ allows for a greater spatial and temporal sampling intensity that improves the \AbbrTDAoI{} and the minimal \AbbrTDAoI{} is achieved with a decrease in sensor distance. In addition, higher spatial correlation helps to improve the \AbbrTDAoI{}, as samples from distant sensors still provide useful information. In this case it is beneficial to increase the intensity of temporal sampling at the expense of the spatial density. Consequently, the optimal sensor distance increases. This is particularly evident in the special case $l_s \to \infty$, in which the sensor distance has no influence at all. Hence, the AeD is zero, and it is best to pool the capacity to serve a single sensor at the highest possible sampling rate to minimize the AoI.
%
%
\subsection{Sensors Connected via Slotted ALOHA Channels}
\label{sec:evaluationaloha}
We investigate the scenario displayed in Fig.~\ref{fig:grid36_2} as before but now with slotted ALOHA channels. The channel model is as follows. The time slot duration is unit sized, e.g., one ms if we choose ms as the unit of time. During a slot, a station either transmits a packet, or it does not perform any operation. The decisions to send (or not) are iid Bernoulli random variables with probability $p$. A transmission is only successful if exactly one station transmits a packet. There are $N$ stations that access the channel and it is well known that the throughput is maximized if $p=1/N$. In this case the probability of successful transmission quickly converges to $1/e$ with increasing $N$. Therefore, the probability that a particular station will successfully transmit in a given time slot is $q=1/(Ne)$. This is the communication budget of the station under the capacity constraint of the channel. The AoI of a sensor connected via the defined slotted ALOHA channel can be easily determined. AoI results for further channel models are available, e.g., for random access policies that are adapted to AoI~\cite{chen:aoirandomaccesschannels} and for non-iid wireless channels~\cite{fidler:ageofinformationparallel}. 

\begin{figure*}
\subfigure[Squared exponential kernel]{\includegraphics[width=0.32\linewidth]{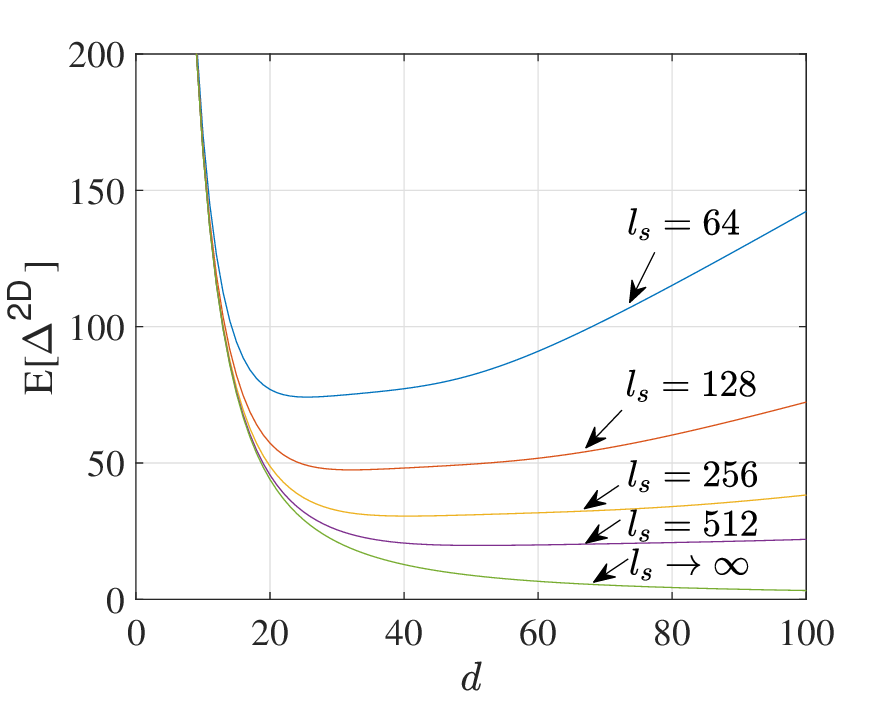}\label{fig:evaluation_AOIvsDistanceALOHA_SE}}
\hfill
\subfigure[Exponential kernel]{\includegraphics[width=0.32\linewidth]{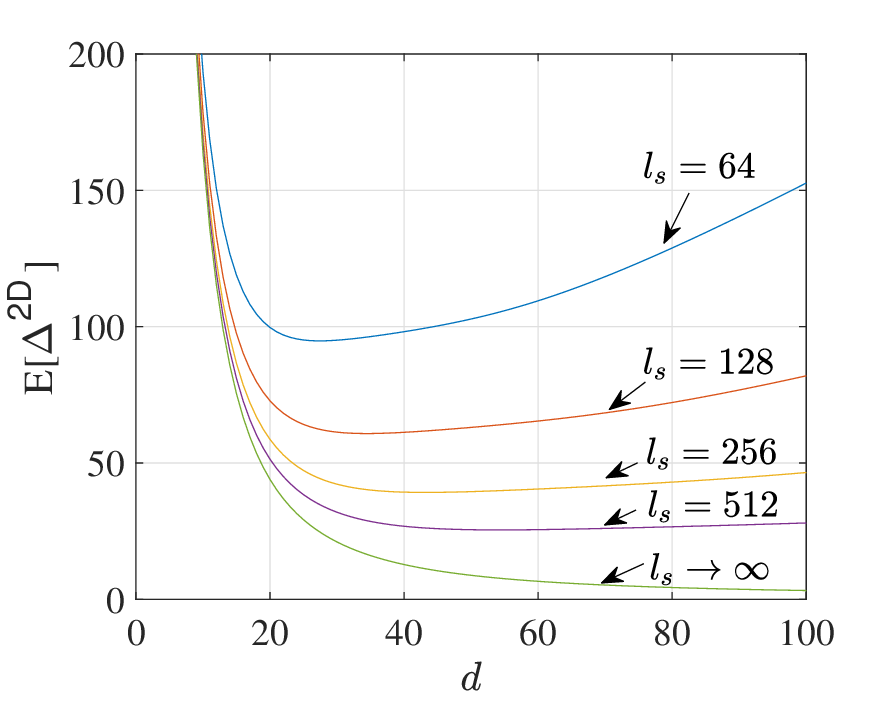}\label{fig:evaluation_AOIvsDistanceALOHA_bound}}
\hfill
\subfigure[Non-independent channels]{\includegraphics[width=0.32\linewidth]{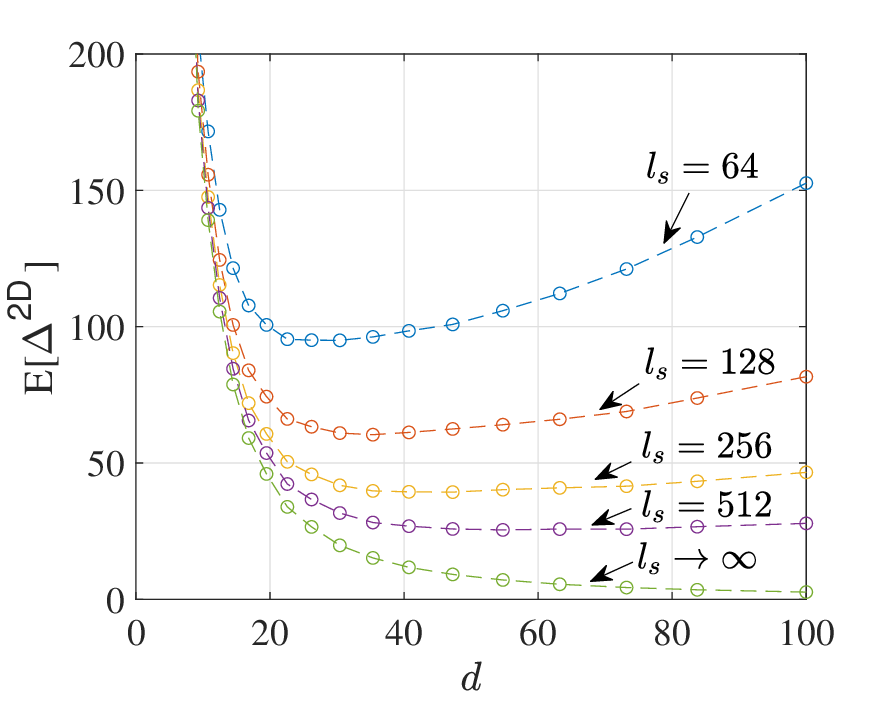}\label{fig:evaluation_AOIvsDistanceALOHA_Sim}}
\caption{Same scenario as in Fig.~\ref{fig:evaluation_AOIvsDistanceMM1}, but with slotted ALOHA channels and squared exponential, shown in Fig.~\ref{fig:evaluation_AOIvsDistanceALOHA_SE}, in addition to exponential kernels in Fig.~\ref{fig:evaluation_AOIvsDistanceALOHA_bound}. For non-independent channels simulation results are shown in Fig.~\ref{fig:evaluation_AOIvsDistanceALOHA_Sim} for the case of exponential kernels.}
\label{fig:evaluation_AOIvsDistanceALOHA}
\end{figure*}

The probability that the AoI of the sensor is larger than a threshold $y$ equals the probability that there was no successful transmission of that sensor during the past $y$ time slots (otherwise the AoI would be less than or at most equal to $y$), resulting in the CCDF
\begin{equation}
\mathsf{P}[\Delta > y] = (1-q)^{\lfloor y \rfloor} ,
\label{eq:ageccdfaloha}
\end{equation}
for  $y \ge 0$. By rounding $y$ down to the nearest integer, the transmission time of one time slot is taken into account in the AoI. Now we consider $S$ distributed sensors, each connected via an independent, homogeneous slotted ALOHA channel with communication budget $q = 1/(Ne)$. We will also be looking at non-independent channels shortly. We show numerical results in Fig.~\ref{fig:evaluation_AOIvsDistanceALOHA}, obtained by substitution of Eq.~\eqref{eq:ageccdfaloha} into Eq.~\eqref{eq:hdaoiexpo} for exponential, Eq.~\eqref{eq:hdaoisquaredexpo} for squared exponential kernels, and Eq.~\eqref{eq:meanspatiotemporalpenalty} for the expected value. 

The results in Fig.~\ref{fig:evaluation_AOIvsDistanceALOHA_bound} are for exponential kernels as also used for Fig.~\ref{fig:evaluation_AOIvsDistanceMM1}. Generally, the \AbbrTDAoI{} for the ALOHA channels that we investigate is higher than in case of the M$\mid$M$\mid$1 queues. This is due to the stricter capacity limitation of the ALOHA channels. Apart from this, the same observations can be made with regard to the sensor distance $d$. When $d$ is small, the communication budget of each sensor is small, resulting in large AoI. When $d$ increases, the influence of the AeD increases. As in Fig.~\ref{fig:evaluation_AOIvsDistanceMM1} the AeD of the tier-1 sensors is $\Lambda = (d/\sqrt{2}) (l_t/l_s)$, e.g., for $d=100$ and $l_s =l_t=128$ it amounts to 71. In between, an optimal $d$ that minimizes the \AbbrTDAoI{} can be identified. 

The minimum in Fig.~\ref{fig:evaluation_AOIvsDistanceALOHA_bound} is less pronounced than in Fig.~\ref{fig:evaluation_AOIvsDistanceMM1}. This is due to the slightly different shape of the AoI of the ALOHA channels compared to the M$\mid$M$\mid$1 queues. As $d$ decreases, each sensor's share of the service rate decreases. In the case of the ALOHA channels, this means that the sensors can access the channel less frequently, but the transmission time of a sample is still one slot. This does not apply to the M$\mid$M$\mid$1 queues, where the transmission time increases as the service rate decreases, further affecting the AoI. One consequence of this is that the tier-2 sensors and above have a somewhat greater effect in the case of the ALOHA channels at small $d$ compared to the case of the M$\mid$M$\mid$1 queues. Numerical results that break down the contribution of the different tiers are shown in Fig.~\ref{fig:evaluation_AOIvsDistanceMM1mu10_tier1_tier2_tier3} in the appendix.

In Fig.~\ref{fig:evaluation_AOIvsDistanceALOHA_SE} we show results for squared exponential kernels with the same parameters and the same ALOHA channel model as in Fig.~\ref{fig:evaluation_AOIvsDistanceALOHA_bound}. The choice of kernel affects the \AbbrTDAoI{} through the AeD, but it does not change the AoI. This means that the differences between Fig.~\ref{fig:evaluation_AOIvsDistanceALOHA_SE} and Fig.~\ref{fig:evaluation_AOIvsDistanceALOHA_bound} are caused by the AeD only. Unlike the AeD of the exponential kernel in Eq.~\eqref{eq:spatialdistanceexpo}, the AeD of the squared exponential kernel in Eq.~\eqref{eq:spatialdistancesquaredexpo} decreases with increasing AoI. We have previously illustrated this effect in Fig.~\ref{fig:quadratickernel}.

When comparing Fig.~\ref{fig:evaluation_AOIvsDistanceALOHA_SE} with Fig.~\ref{fig:evaluation_AOIvsDistanceALOHA_bound} the differences are most visible in the middle part for moderate $d$. In contrast, when $d$ is small, the \AbbrTDAoI{} is dominated by the AoI which is independent of the kernel, and when $d$ is large, the AoI becomes small and the AeD of the squared exponential kernel and the exponential kernel converge (in the limit for $\Delta_t(\varsigma)=0$, Eq.~\eqref{eq:spatialdistancesquaredexpo} becomes identical to Eq.~\eqref{eq:spatialdistanceexpo}). In the middle part close to the value of $d$ that minimizes the \AbbrTDAoI{} both the AeD and the AoI are significant. In this case, samples from more distant sensors can make an important contribution to improving the \AbbrTDAoI{}, when fresh samples from nearby sensors are not available, in particular for the squared exponential kernel, where the AeD decreases with increasing AoI.

In addition to independent ALOHA channels, we have also investigated non-independent channels. Since Eq.~\eqref{eq:spatiotemporaldistanceccdflinear} is not applicable in this case, we resort to simulations. In the simulation, each of the $N$ sensors per area generates and transmits a sample in each time slot with probability $p=1/N$. If two or more sensors transmit a sample in the same time slot, there is a collision and the samples are lost. For large $N$, the throughput of the channel converges to $1/e$ as in the independent channel model used before. We provide simulation results for non-independent channels in Fig.~\ref{fig:evaluation_AOIvsDistanceALOHA_Sim} for the case of exponential kernels that we considered also in Fig.~\ref{fig:evaluation_AOIvsDistanceALOHA_bound}. We have also produced simulation results for the case of independent channels. We do not show these as they exactly match the numerical results of the analysis in Fig.~\ref{fig:evaluation_AOIvsDistanceALOHA_bound}. Comparing Fig.~\ref{fig:evaluation_AOIvsDistanceALOHA_Sim} with Fig.~\ref{fig:evaluation_AOIvsDistanceALOHA_bound} we see that the results for non-independent and independent channels are only slightly different. The differences arise from the fact that in the case of independent channels, two or more samples can be successfully transmitted in the same time slot (of which the sample with the lowest \AbbrTDAoI{} is selected), while in the case of the non-independent channel, a collision occurs and the samples are lost. In our simulations, however, the effect of this difference is negligible.
%
%
\subsection{Prediction Variance}
\label{sec:evaluationpredictionvariance}
We complement our results by looking at the mean posterior variance $\mathsf{E}[\Phi]$ when predicting the process at the center (black dot) of Fig.~\ref{fig:grid36_2} from the samples that are available at the monitor from the $S=16$ closest sensors (green dots). The prediction variance is directly related to the \AbbrTDAoI{} via Eq.~\eqref{eq:predictionvariancesinglesamplesolved}, which expresses $\Phi$ as a function of $\TDAoI$ using the temporal correlation $g$. 
\begin{figure*}
\subfigure[Squared exponential kernel]{\includegraphics[width=0.32\linewidth]{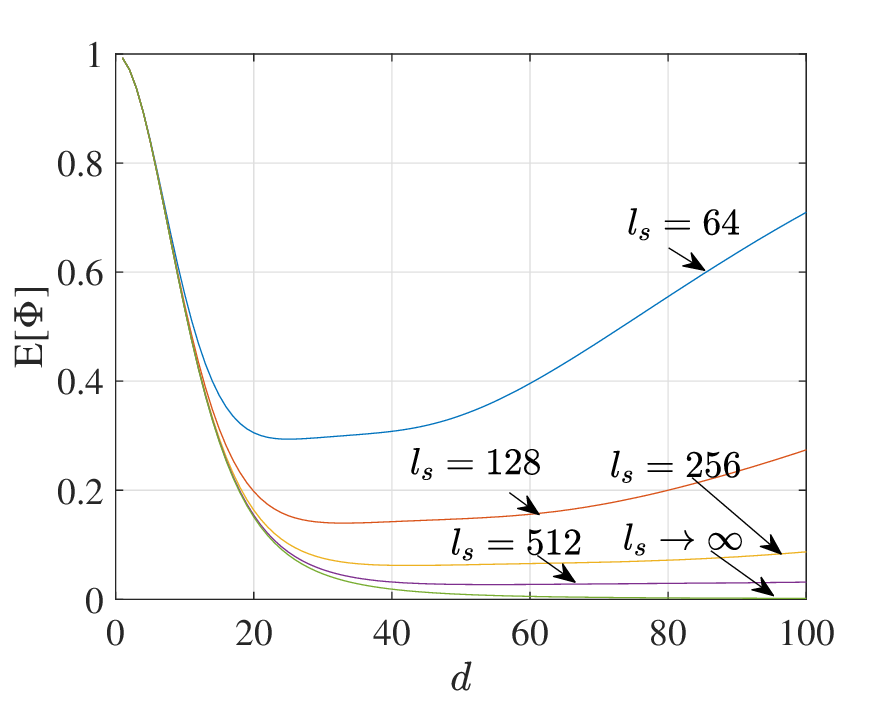}\label{fig:evaluation_VariancevsDistanceALOHA_SE}}
\hfill
\subfigure[Exponential kernel]{\includegraphics[width=0.32\linewidth]{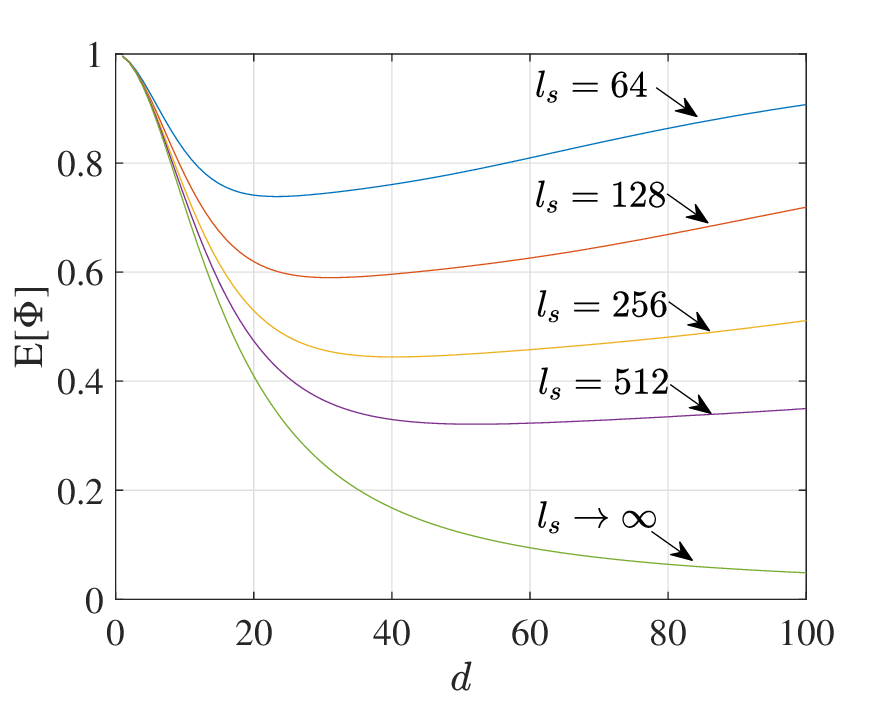}\label{fig:evaluation_VariancevsDistanceALOHA_LE}}
\hfill
\subfigure[Best (solid) vs. all samples (dashed)]{\includegraphics[width=0.32\linewidth]{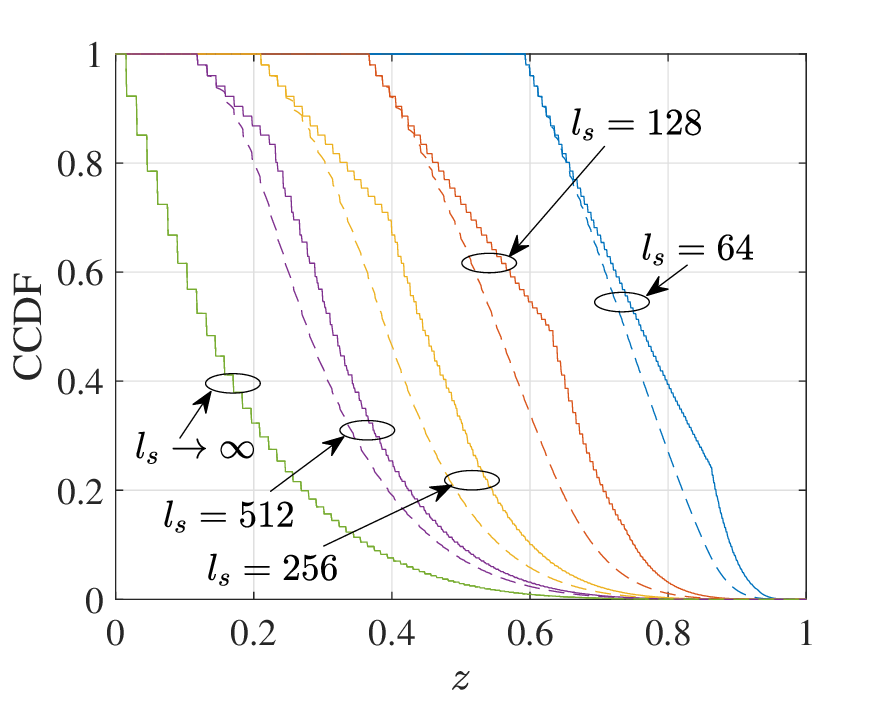}\label{fig:evaluation_bestvsallsamples}}
\caption{Figs.~\ref{fig:evaluation_VariancevsDistanceALOHA_SE} and~\ref{fig:evaluation_VariancevsDistanceALOHA_LE} show the mean prediction variance for the scenario evaluated in Figs.~\ref{fig:evaluation_AOIvsDistanceALOHA_SE} and~\ref{fig:evaluation_AOIvsDistanceALOHA_bound} if the best sample is used for prediction. Fig.~\ref{fig:evaluation_bestvsallsamples} shows CCDFs of the prediction variance for the exponential kernel, $d=40$, and the case where either the best sample (solid lines) or all samples that are available at the monitor (dashed lines) are used for prediction.}
\label{fig:evaluation_VariancevsDistanceALOHA}
\end{figure*}

In general, Eq.~\eqref{eq:predictionvariance} can take all samples $\Gamma_t(\varsigma,\tau)$ that are available at the monitor at time $t$ to make the best prediction. We can derive an analytical estimate of the prediction variance if we use only the most recent sample of the sensor $\varsigma$ that minimizes the prediction variance $\mathrm{arg\,min}_{\varsigma} \{ \Phi_t(\varsigma,s) \}$. This is the sample that has the minimal \AbbrTDAoI{}. In the case that the sensors are connected via independent channels Eq.~\eqref{eq:spatiotemporaldistanceccdfgeneral} applies accordingly to the prediction variance and the CCDF of the minimal prediction variance $\Phi_t(s) = \min_{\varsigma} \{ \Phi_t(\varsigma,s) \}$ is
\begin{align}
\mathsf{P} [\Phi_t(s) > z\sigma^2 ] &= \mathsf{P} \Bigl[\min_{\varsigma \in \mathbb{S}} \{ \Phi_t(\varsigma,s) \} > z\sigma^2 \Bigr] \nonumber \\ 
&= \prod_{\varsigma \in \mathbb{S}} \mathsf{P} [ \Phi_t(\varsigma,s) > z\sigma^2 ] . 
\label{eq:predictionvarianceindependence}
\end{align}
Depending on the correlation kernel, exponential, squared exponential, or rational quadratic, we can substitute~\eqref{eq:predvarexpo}, \eqref{eq:predvarsquaredexpo}, or~\eqref{eq:predvarrationalquadratic} for $\Phi_t(\varsigma,s)$ into Eq.~\eqref{eq:predictionvarianceindependence}. We then obtain the mean prediction variance by integrating the CCDF as $\mathsf{E}[\Phi(s)] = \int_0^1 \mathsf{P} [\Phi(s) > z \sigma^2] dz$.

We consider the case where the sensors are connected via independent slotted ALOHA channels as in Sec.~\ref{sec:evaluationaloha}. The mean prediction variance is shown in Figs.~\ref{fig:evaluation_VariancevsDistanceALOHA_SE} and~\ref{fig:evaluation_VariancevsDistanceALOHA_LE} that correspond to the \AbbrTDAoI{} shown in Figs.~\ref{fig:evaluation_AOIvsDistanceALOHA_SE} and~\ref{fig:evaluation_AOIvsDistanceALOHA_bound}. While the mean prediction variance follows the general trend of the \AbbrTDAoI{}, the effect of the two different temporal correlation functions $g$ (squared exponential and exponential, respectively) on the mean prediction variance is clearly visible.

Fig.~\ref{fig:evaluation_bestvsallsamples} also shows the CCDFs of the prediction variance for the case of exponential kernels as in Fig.~\ref{fig:evaluation_VariancevsDistanceALOHA_LE} and parameter $d=40$. In this experiment, we also include the posterior variance that is computed in a simulation by evaluating Eq.~\eqref{eq:predictionvariance} using all samples that the monitor collected during the past $T$ time slots (dashed lines), compared to using only the best sample in Eq.~\eqref{eq:predictionvarianceindependence} (solid lines). For computational reasons, we restrict $T=1000$. In the limit for infinite spatial correlation $l_s \to \infty$, all sensors observe the same process and only the most recent sample is relevant. Hence, using more samples does not show an effect. For finite $l_s$, more samples generally improve the prediction. This becomes visible at the bends of the solid curves. These bends occur when samples of different sensors become equally valuable. This happens, e.g., when the \AbbrTDAoI{} of nearby sensors exceeds the AeD of more distant sensors, so that the samples of distant sensors can potentially be equally good or better. We illustrated the effect in Fig.~\ref{fig:serviceprovisioning}. Here, using all samples together (dashed curves) shows an improvement compared to using only the best sample (solid curves). The mean values that correspond to the CCDFs in Fig.~\ref{fig:evaluation_bestvsallsamples} are summarized in Tab.~\ref{tab:expectedvalues}. Using all samples gives the lowest mean prediction variance, whereby the best sample with the lowest \AbbrTDAoI{} is the dominant factor and achieves comparably good results on its own.
\begin{table}
\centering
\caption{Mean values for Fig.~\ref{fig:evaluation_bestvsallsamples}.}
\begin{tabular}{r|ccccc} 
$\mathsf{E}[\Phi]$ & $l_s = 64$ & $l_s = 128$ & $l_s = 256$ & $l_s = 512$ & $l_s \to \infty$ \\ \hline
best sample & 0.761 & 0.597 & 0.445 & 0.330 & 0.169 \\
all samples & 0.738 & 0.553 & 0.402 & 0.301 & 0.169 
\end{tabular}
\label{tab:expectedvalues}
\end{table}
%
%
\section{Conclusions}
\label{sec:conclusions}
This work presented a model of AoI for two-dimensional processes. The physical phenomenon can be a spatio-temporal process observed by distributed sensors. Alternatively, in the case of inhomogeneous sensors, the type of sensor can be considered as an additional dimension. By employing a Gaussian process model we have shown how the spatial distance between sensors can be transformed into an Age-equivalent Distance (AeD). The AeD can be a constant offset that does not change over time, depending on the correlation kernel, or it can diminish as time progresses. This means that a sample from a distant sensor becomes increasingly valuable when the AoI of a nearby sensor is high. By transforming distance into the domain of AoI, one can leverage a variety of established results for AoI and AoI of parallel systems. We presented evaluation results for different models of communications links, different network topologies, and different kernel functions. These make it possible to optimize sensor density, to decide on the necessity for additional sensors, and to configure the network capacity to ensure that desired performance criteria are met. The \AbbrTDAoI{} model can be instantiated in different ways extending the utility of the established theory to a range of potential applications.
%
%
\balance
\bibliographystyle{IEEEtran}
\bibliography{IEEEabrv,IEEEfidler}
%
%
\section{Appendix}
%
%
\subsection{Background on Gaussian Process Regression}\label{appendix:GP}
\label{sec:gaussianprocesses}
The following Gaussian regression model is commonly used for prediction in machine learning applications~\cite{rasmussen:gaussianprocesses}. For an intuitive introduction see, e.g.,~\cite{turner:gaussianprocesses}. 
%
%
\subsubsection{Process Definition}
We consider a Gaussian process $f({\bf x})$ that takes an input vector ${\bf x}$ to a function value $f$. The Gaussian process is completely specified by its mean function $m({\bf x}) = \mathsf{E}[f({\bf x})]$ and covariance function $k({\bf x},{\bf x'}) = \mathsf{E}[(f({\bf x}) - m({\bf x})) (f({\bf x'}) - m({\bf x'}))]$. 

The covariance function is also called the kernel and there are a number of well-known kernel functions in the literature~\cite{Duvenaud:kernelcookbook} that are characteristic of different processes. For a first example we consider an exponential covariance function of a scalar input $x$, that is the case of the Ornstein-Uhlenbeck process~\cite{rasmussen:gaussianprocesses}. It is defined as
\begin{equation*}
k(x,x') = \sigma^2 e^{-\frac{1}{l} \mid x-x' \mid},
\end{equation*}
where $\sigma^2 > 0$ is the variance and parameter $l > 0$ determines the length scale of the correlation. The kernel depends only on the difference $x-x'$, i.e., it is stationary. The mean function is frequently normalized as $m({\bf x}) = 0$.

For a set of values $\{ ({\bf x}_i,f({\bf x}_i), i=1,\dots,m \}$ the Gaussian process implies a multivariate Gaussian distribution ${\bf f} = \{f({\bf x}_1), f({\bf x}_2), \dots, f({\bf x}_m)\} \sim \mathcal{N}({\pmb \mu}_{\bf f},\Sigma_{\bf ff})$ where ${\pmb \mu}_{\bf f}$ is the mean vector and $\Sigma_{\bf ff}$ the covariance matrix of ${\bf f}$ obtained by evaluating the mean function and covariance function at all points $\{{\bf x}_1, {\bf x}_2, \dots {\bf x}_m\}$. 

In practice, the observation of the function values $f({\bf x}_i)$ may be subject to noise. It is straight-forward to add independent Gaussian observation noise to the covariance matrix $\Sigma_{\bf ff}$~\cite[Eq. (2.20)]{rasmussen:gaussianprocesses}.
%
%
\subsubsection{Regression/Prediction}
Given a set of $m$ observed values ${\bf f}$, the goal is to predict a set of $m_*$ values ${\bf f}_*$ that are not observed. The joint prior distribution of ${\bf f}$ and ${\bf f}_*$ is
\begin{equation*}
\begin{bmatrix}
{\bf f} \\           
{\bf f}_* 
\end{bmatrix}
\sim
\mathcal{N}
\begin{pmatrix}  
\begin{bmatrix}
{\pmb \mu}_{\bf f} \\           
{\pmb \mu}_{{\bf f}_*} 
\end{bmatrix},
\begin{bmatrix}
\Sigma_{\bf ff} & \Sigma_{{\bf ff}_*} \\
\Sigma_{{\bf f}_*{\bf f}} & \Sigma_{{\bf f}_* {\bf f}_*}
\end{bmatrix}
\end{pmatrix},
\end{equation*}
where $\Sigma_{{\bf ff}_*}$ is the $m \times m_*$ covariance matrix of all pairs of values in ${\bf f}$ and ${\bf f}_*$ and so on. Conditioning on the observations ${\bf f}$, the posterior distribution of ${\bf f}_*$ is
\begin{align}
[{\bf f}_* \mid {\bf f}] \sim \mathcal{N} (&{\pmb \mu}_{{\bf f}_*} + \Sigma_{{\bf f}_*{\bf f}} \Sigma^{-1}_{{\bf f}{\bf f}} ({\bf f} - {\pmb \mu}_{\bf f}) , \nonumber \\ & \Sigma_{{\bf f}_*{\bf f}_*} - \Sigma_{{\bf f}_*{\bf f}} \Sigma^{-1}_{{\bf f}{\bf f}} \Sigma_{{\bf f}{\bf f}_*} ) .
\nonumber 
\end{align}
It is noted that the posterior covariance does not depend on the actual values that are observed. The term $\Sigma_{{\bf f}_*{\bf f}} \Sigma^{-1}_{{\bf f}{\bf f}} \Sigma_{{\bf f}{\bf f}_*}$ quantifies the reduction in uncertainty that is achieved by the observations ${\bf f}$.
%
%
\subsubsection{Multi-dimensional Kernels}
The choice of the process covariance function, i.e., the kernel, is a much-considered question. In brief, the kernel models the degree of similarity between data points. The kernel can be estimated from training data or it can be defined by a model. Common models are exponential, squared exponential, and rational quadratic.

If the process depends on a vector of inputs, several individual kernels can be combined to build a new multi-dimensional kernel. Multiplication of kernels gives a kernel that has large values only, if all of the individual kernels have a large value. This can be thought of as \lq{}and\rq{} operation~\cite{Duvenaud:kernelcookbook}. In this work, we examine a system of spatially distributed sensors whose sensor values are correlated in both space and time. Other higher-dimensional models are also a possibility.

In the event that the spatial and temporal correlation are each modeled by an exponential covariance function, with index $s$ and $t$, respectively, the two-dimensional kernel that is obtained by multiplication is
\begin{equation*}
k({\bf x}, {\bf x'}) = \sigma^2 e^{-\frac{1}{l_s} \mid x_s-x_s' \mid - \frac{1}{l_t} \mid x_t-x_t' \mid} .
\end{equation*}
%
%
\subsection{Rational Quadratic Kernels}\label{appendix:rational_quadratic_kern}
We also instantiate the \AbbrTDAoI{} model of Gaussian processes derived in Sec.~\ref{sec:hdaoi} for rational quadratic product kernels. We have $g(\Delta_t(\varsigma)) = (1 + \Delta_t(\varsigma)^2/(2\beta l_t^2) )^{-\beta}$ and $h(\varsigma,s) = (1 + (x_{\varsigma}-x_s)^2/(2\alpha l_s^2) )^{-\alpha}$ so that 
\begin{equation*}
g^{-1}(y) = \sqrt{2 \beta l_t^2 \bigl(y^{-\frac{1}{\beta}} -1\bigr)} ,
\end{equation*}
for $y \in (0,1]$. By insertion into Eq.~\eqref{eq:lambda} the AeD is
\begin{equation}
\Lambda_t(\varsigma,s) = \sqrt{ 2 \beta l_t^2 \bigl(h(\varsigma,s)^{-\frac{1}{\beta}} - 1 \bigr) + h(\varsigma,s)^{-\frac{1}{\beta}} \Delta_t(\varsigma)^2 } - \Delta_t(\varsigma) .
\label{eq:spatialdistancerationalquadratic}
\end{equation}
The CCDF of the \AbbrTDAoI{} is obtained from Eq.~\eqref{eq:spatiotemporaldistanceccdf} as
\begin{equation*}
\mathsf{P}[\TDAoI_t(\varsigma,s) > y] = \mathsf{P} \biggl[ \Delta_t(\varsigma) > \Bigl(h(\varsigma,s)^{\frac{1}{\beta}} \bigl( y^2 + 2 \beta l_t^2 \bigr) - 2 \beta l_t^2\Bigr)^{\frac{1}{2}} \biggr]  ,
\end{equation*}
for $y \ge (2\beta l_t^2 (h(\varsigma,s)^{-\frac{1}{\beta}}-1))^{\frac{1}{2}}$, and the CCDF of the prediction variance Eq.~\eqref{eq:predictionvarianceccdf} becomes
\begin{equation}
\mathsf{P} [ \Phi_t(\varsigma,s) > z \sigma^2 ] = \mathsf{P} \left[ \Delta_t(\varsigma) > \Bigl(2 \beta l_t^2 \bigl((1-z)^{-\frac{1}{2\beta}}  h(\varsigma,s)^{\frac{1}{\beta}} -1\bigr)\Bigr)^{\frac{1}{2}}\right] , 
\label{eq:predvarrationalquadratic}
\end{equation}
for $z \ge 1-h(\varsigma,s)^2$ and $z < 1$. 

The AeD of the rational quadratic kernel in Eq.~\eqref{eq:spatialdistancerationalquadratic} exhibits a similar behavior with respect to $\Delta_t(\varsigma)$ as the AeD of the squared exponential kernel in Eq.~\eqref{eq:spatialdistancesquaredexpo}. For high $\alpha$ and $\beta$ the rational quadratic kernel converges against the squared exponential kernel. Consequently letting $\alpha \to \infty$ and $\beta \to \infty$ the equations for $\Lambda$, $\mathsf{P}[\TDAoI_t(\varsigma,s) > y]$ and $\mathsf{P} [ \Phi_t(\varsigma,s) > z \sigma^2 ]$ are verified to be the same as for the squared exponential kernel.
%
%
\subsection{\AbbrTDAoI{} of Mixed Product Kernels}\label{appendix:mixed_product_kern}
A variety of other kernel functions are possible, such as any product combinations of the kernels used so far, e.g., a product of an exponential kernel for temporal and a squared exponential kernel for spatial correlation, etc.

In all cases where the temporal correlation has an exponential kernel $g(\Delta_t(\varsigma)) = e^{-\frac{1}{l_t} \Delta_t(\varsigma)}$ we generally have an AeD of the structure
\begin{equation*}
\Lambda(\varsigma,s) = -l_t \ln h(\varsigma,s) ,    
\end{equation*}
and $\Lambda(\varsigma,s)$ is a function of space only and is independent of time. Conversely, for the squared exponential temporal kernel $g(\Delta_t(\varsigma)) = e^{-\frac{1}{2l_t^2} \Delta_t(\varsigma)^2}$ the AeD
\begin{equation*}
\Lambda_t(\varsigma,s) = \sqrt{-2  l_t^2 \ln h(\varsigma,s) + \Delta_t(\varsigma)^2} - \Delta_t(\varsigma)
\end{equation*}
generally takes a structure similar to Eq.~\eqref{eq:spatialdistancesquaredexpo} that depends on space and time. 
%
%
\subsection{\AbbrTDAoI{} with a Random AeD}\label{appendix:random_AeD}
In this section, we show how a random AeD can be included in the \AbbrTDAoI{} model. The application scenario is based on~\cite{popovski2019multiplesensors, popovski2023multiplesensors, modiano:aoicorrelatedsources}, where distributed sensors can detect the state of several sources, but only with a certain probability. 

In the scenario that we investigate, there is a set of distributed but overlapping sensors $\mathbb{S}$, such as cameras. Each of the sensors $\varsigma \in \mathbb{S}$ can take samples of the physical process at its own position $x_{\varsigma}$ at will. Due to the overlap, sensor $\varsigma$ can also generate samples of the physical process at other positions $x_s$, but depending on its distance, the sensor will only be successful with probability $p_{\varsigma s} \in [0,1]$. The sensor is aware of its success and either retains or discards the sample accordingly. For instance, a camera tracking an object can also detect if it lost sight of the object, e.g. due to occlusion. In view of the limited reliability, the generation of samples for other positions $x_s \neq x_{\varsigma}$ is carried out as a secondary task. We model this as an independent sampling process with rate $\omega_{\varsigma}$.

Whenever a sensor $\varsigma$ sends a fresh sample of the physical process at its own position $x_{\varsigma}$, it also adds the most recent successful samples that it has obtained for the other positions $x_s \neq x_{\varsigma}$. In this model, the sensor sends fresh information for its own position but potentially aged information for the other positions. We consider this as the AeD of sensor $\varsigma$ with respect to position $x_s$ to determine the \AbbrTDAoI{}. Unlike in~\cite{popovski2019multiplesensors, popovski2023multiplesensors, modiano:aoicorrelatedsources} we do not consider scheduled transmissions but the sensors transmit their samples via independent slotted ALOHA channels as defined in Sec.~\ref{sec:evaluationaloha}. 

An example of the selected scenario is in vehicular communications. The physical process that we consider is the position $x_s$ of the vehicle $s$ itself. The vehicle $s$ can sample its internal positioning sensors at will and from time to time it broadcasts its position via so-called coope\-ra\-tive awareness messages. This information ages over time and messages can also be lost. The other vehicles $\varsigma$ use their sensors, such as camera, lidar, or radar, to detect their surroundings and, among other things, to track vehicle $s$. The detection is not reliable, e.g. due to occlusions. In their broadcasts, the vehicles also indicate the estimated positions of the detected objects, including vehicle $s$. By fusing these data, a collective perception service is achieved.

We focus on one position $x_s$ and determine the \AbbrTDAoI{} of sensor $\varsigma \neq s$ with respect to this position. First, we consider the case where sensor $\varsigma$ has perfect detection of the physical process at position $x_s$, i.e. $p_{\varsigma s}=1$. We model the sampling process as a Poisson process with rate $\omega_{\varsigma}$. This implies that the times between consecutive samples are iid exponential random variables with parameter $\omega_{\varsigma}$. If we observe the sampling process at a random point in time $t$, the time since the last sample, that is the backward recurrence time, is exponential with parameter $\omega_{\varsigma}$~\cite[Ex. 5.5]{haverkort:performancebook}. Other sampling processes are also possible.

If $p_{\varsigma s}<1$, we have to consider the case of unsuccessful samples, which occur with probability $1-p_{\varsigma s}$. Hence the time between two successful samples $Y$ becomes $Y = \sum_{i=1}^{N} X_i$. Here $X_i$ are the iid exponential inter-arrival times of the sampling process with parameter $\omega_{\varsigma}$ and given the successes are iid, $N$ is a geometric random variable with $\mathsf{P}[N=n] = p_{\varsigma s} (1-p_{\varsigma s})^{n-1}$ for $n \in \{1,2,\dots\}$. It follows that $Y$ is exponential with parameter $p_{\varsigma s} \omega_{\varsigma}$. This is seen by unconditioning the conditional moment generating function $\mathsf{M}[Y|N=n] = (\omega_{\varsigma}/(\omega_{\varsigma}-\theta))^n$ which gives 
\begin{align*}
\mathsf{M}[Y] &= \sum_{n=1}^{\infty} \Bigl(\frac{\omega_{\varsigma}}{\omega_{\varsigma}-\theta}\Bigr)^n p_{\varsigma s} (1-p_{\varsigma s})^{n-1} \\
&= \frac{p_{\varsigma s}\omega_{\varsigma}}{p_{\varsigma s}\omega_{\varsigma}-\theta} ,
\end{align*}
for $\theta < p_{\varsigma s} \omega_{\varsigma}$. The result is the moment generating function of the exponential distribution with parameter $p_{\varsigma s} \omega_{\varsigma}$ and thus the sequence of successful samples is a Poisson process. If this process is observed at a random point in time, e.g. at the time when the sensor $\varsigma$ sends its samples, the time since the sensor $\varsigma$ last successfully sampled the position $x_s$ is again exponential with parameter $p_{\varsigma s} \omega_{\varsigma}$. This is the AeD of sensor $\varsigma$ with respect to position $x_s$.

For $\varsigma \neq s$, we derive the CCDF of the \AbbrTDAoI{} from Eq.~\eqref{eq:spatialdistance} by convolution of the CDF of the AoI 
of the slotted ALOHA channel given by Eq.~\eqref{eq:ageccdfaloha} with the probability density function of the AeD that is $\text{pdf}(x) = p_{\varsigma s}\omega_{\varsigma} e^{- p_{\varsigma s} \omega_{\varsigma} x}$ as
\begin{align}
\mathsf{P}[\TDAoI_t(\varsigma,s) > y] &= 1-\mathsf{P}[\Delta_t(s) + \Lambda_t(\varsigma,s) \le y] \nonumber \\
&= 1 - \int_0^y \bigl(1-(1-q)^{y-x}\bigr) p_{\varsigma s} \omega_{\varsigma} e^{- p_{\varsigma s} \omega_{\varsigma} x} dx \nonumber \\
&= \frac{ \ln(1-q) e^{-p_{\varsigma s} \omega_{\varsigma} y} + p_{\varsigma s} \omega_{\varsigma} (1-q)^y }{\ln(1-q) + p_{\varsigma s} \omega_{\varsigma}} ,
\label{eq:hdaoipopovskimodel}
\end{align}
where we used an estimate of Eq.~\eqref{eq:ageccdfaloha} without the integer constraint. For $\varsigma = s$ we have $\TDAoI_t(s,s) = \Delta_t(s)$ and $\mathsf{P}[\Delta(s) > y] = (1-q)^y$. The minimal \AbbrTDAoI{} of several sensors $\varsigma \in \mathbb{S}$ follows by insertion of Eq.~\eqref{eq:hdaoipopovskimodel} into Eq.~\eqref{eq:spatiotemporaldistanceccdfgeneral}. 

The \AbbrTDAoI{} in Eq.~\eqref{eq:hdaoipopovskimodel} decreases with increasing $p_{\varsigma s}\omega_{\varsigma} > 0$, 
and in the limit for $p_{\varsigma s}\omega_{\varsigma} \to \infty$ we have $\mathsf{P}[\TDAoI_t(\varsigma,s) > y] = (1-q)^y$. For $p_{\varsigma s}\omega_{\varsigma} < -\ln(1-q)$, increasing $p_{\varsigma s}\omega_{\varsigma}$ significantly improves the \AbbrTDAoI{}. The gain becomes saturated when $p_{\varsigma s}\omega_{\varsigma} > -\ln(1-q)$. The reason for this is that the tail decay rate of Eq.~\eqref{eq:hdaoipopovskimodel} is either dominated by the term $e^{-p_{\varsigma s} \omega_{\varsigma} y}$ or by the term $(1-q)^y=e^{\ln(1-q) y}$ depending on which value decays slower. If $p_{\varsigma s}\omega_{\varsigma} > -\ln(1-q)$ the slotted ALOHA channel is the limiting factor and increasing the rate of successful samples has little effect. In practice, technical constraints will also limit the sampling rate, and a high sampling rate can affect the probability of success $p_{\varsigma s}$ and the independence of sampling successes.
%
%
\subsection{Breakdown of Contribution by Sensor Tiers}\label{appendix:sensor_tiers}
\begin{figure}
\centering
\hfill
\subfigure[M$\mid$M$\mid$1]{\includegraphics[width=0.41\linewidth]{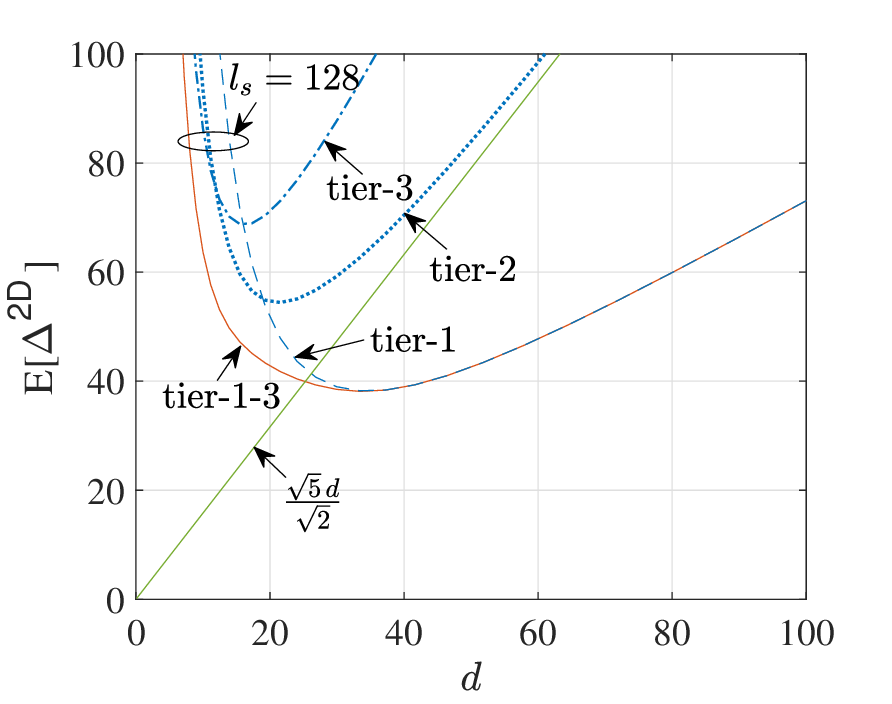}\label{fig:MM1_ls128_s36_tier1_tier2_tier3}}
\hfill
\subfigure[Slotted ALOHA]{\includegraphics[width=0.41\linewidth]{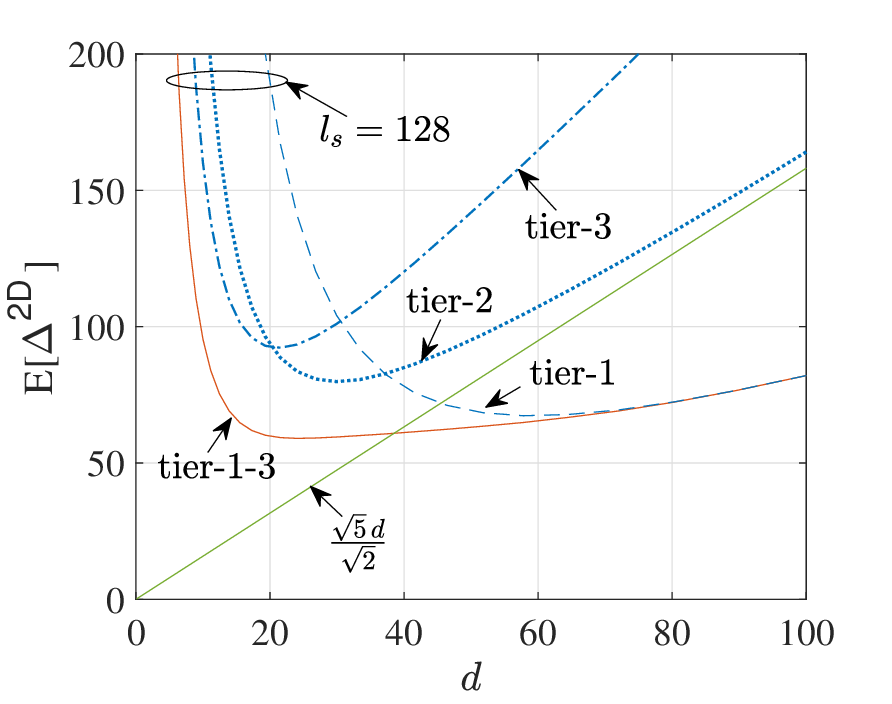}\label{fig:ALOHA_ls128_s36_tier1_tier2_tier3}}
\hfill
\caption{The samples of $S=36$ sensors are used by the monitor compared to $S=16$ in Sec.~\ref{sec:evaluation}. Fig.~\ref{fig:MM1_ls128_s36_tier1_tier2_tier3} is for M$\mid$M$\mid$1 queues as in Fig.~\ref{fig:evaluation_AOIvsDistanceMM1mu10} where $\mu=10$ and $l_s=128$, and Fig.~\ref{fig:ALOHA_ls128_s36_tier1_tier2_tier3} for slotted ALOHA channels as in Fig.~\ref{fig:evaluation_AOIvsDistanceALOHA_bound} for $l_s=128$. The results are divided according to the contribution of tier-1, tier-2, and tier-3 sensors. The green line denotes the AeD of the closest tier-2 sensors.}
\label{fig:evaluation_AOIvsDistanceMM1mu10_tier1_tier2_tier3}
\end{figure}

We consider the sensor grid displayed in Fig.~\ref{fig:grid36_2} and defined in Sec.~\ref{sec:evaluation} and break down the contributions of the different sensor tiers. Compared to $S=16$ tier-1 and 2 sensors in Sec.~\ref{sec:evaluation}, we now consider the case of $S=36$ tier-1, 2, and 3 sensors, where tier-1 comprises a square of the closest 4 sensors, tier-2 the next 12, and tier-3 the next 20 sensors. The results are shown in Fig.~\ref{fig:evaluation_AOIvsDistanceMM1mu10_tier1_tier2_tier3}, where  Fig.~\ref{fig:MM1_ls128_s36_tier1_tier2_tier3} is for the case of M$\mid$M$\mid$1 queues and corresponds to Fig.~\ref{fig:evaluation_AOIvsDistanceMM1mu10} for $\mu=10$ and $l_s=128$, and Fig.~\ref{fig:ALOHA_ls128_s36_tier1_tier2_tier3} is for the slotted ALOHA channels as in Fig.~\ref{fig:evaluation_AOIvsDistanceALOHA_bound} also for $l_s=128$.

We start with the case of the M$\mid$M$\mid$1 queues shown in Fig.~\ref{fig:MM1_ls128_s36_tier1_tier2_tier3}, where the \AbbrTDAoI{} when using all $S=36$ tier-1, 2, and 3 sensors together is shown by the red solid line. The minimal \AbbrTDAoI{} at $d=34$ is almost exclusively due to the tier-1 sensors, whose \AbbrTDAoI{} is marked by the blue dashed line. The closest tier-2 sensors have an AeD of $\sqrt{5} d/\sqrt{2}$ marked by the green line. With increasing $d$, the dotted blue line, that is the \AbbrTDAoI{} of only the tier-2 sensors, approaches the AeD and the tier-2 sensors do not bring any improvement when used in addition to the tier-1 sensors. The tier-2 sensors contribute to improving the \AbbrTDAoI{} left of the minimum. The effect becomes noticeable for $d < 25.3$, when the AeD of the tier-2 sensors becomes smaller than the mean \AbbrTDAoI{} of 40 that is achieved by the tier-1, 2, and 3 sensors together. With decreasing $d$ the contribution of the tier-2 sensors eventually outweighs that of the tier-1 sensors at $d\le18.9$, which is a consequence of the higher number of 12 tier-2 sensors compared to 4 tier-1 sensors. In this region the inclusion of the tier-3 sensors also starts to provide a slight advantage. We omit showing graphs for the case $\mu=100$ and $\mu=1000$ as in Figs.~\ref{fig:evaluation_AOIvsDistanceMM1mu100} and~\ref{fig:evaluation_AOIvsDistanceMM1mu1000} since they show the same behavior.  

In case of the slotted ALOHA channels in Fig.~\ref{fig:ALOHA_ls128_s36_tier1_tier2_tier3} the influence of the higher tiers is somewhat larger. The minimum of the \AbbrTDAoI{} curve achieved by the tier-1, 2, and 3 sensors together is less pronounced and the contribution of the AoI of the ALOHA channel is larger than for the M$\mid$M$\mid$1 queues (reflected in the different scaling of the y-axis). Due to the larger AoI, sensors with a larger AeD can still be useful, as seen for the tier-2 and tier-3 sensors, which all contribute to the minimal mean \AbbrTDAoI{} that is obtained at $d=24$. In comparison, with $S=16$ sensors in Fig.~\ref{fig:evaluation_AOIvsDistanceALOHA_bound}, the minimal mean \AbbrTDAoI{} is obtained at $d=34.5$. The difference of the minimal mean \AbbrTDAoI{} of 59 for $S=36$ sensors in Fig.~\ref{fig:ALOHA_ls128_s36_tier1_tier2_tier3} and 60.8 for $S=16$ sensors in Fig.~\ref{fig:evaluation_AOIvsDistanceALOHA_bound} is, however, small.
%
%
\end{document}